\documentclass[pdftex,twocolumn,epjc3]{svjour3}
\RequirePackage[T1]{fontenc}
\smartqed  
\RequirePackage{graphicx}
\RequirePackage{mathptmx}      
\RequirePackage{flushend}
\RequirePackage[numbers,sort&compress]{natbib}
\RequirePackage[colorlinks,citecolor=blue,urlcolor=blue,linkcolor=blue]{hyperref}

\usepackage{amsmath,mathtools}
\usepackage{upgreek}
\usepackage{cancel}
\usepackage{subcaption}
\usepackage{datetime}
\usepackage{lipsum}
\usepackage{orcidlink}

\newcommand{\D}{\mathrm{d}}
\newcommand{\CF}{C_\mathrm{F}}
\newcommand{\CA}{C_\mathrm{A}}

\newcommand{\TR}{T_\mathrm{R}}
\newcommand{\alphaS}{\alpha_\mathrm{s}}
\newcommand{\MUF}{\mu_\mathrm{F}}

\newcommand{\MSBAR}{\overline{\mathrm{MS}}}

\newcommand{\pythia}{\textsc{Pythia}}
\newcommand{\sherpa}{\textsc{Sherpa}}
\newcommand{\powhegbox}{\textsc{Powheg\,Box}}
\newcommand{\vincia}{\textsc{Vincia}}

\journalname{Eur. Phys. J. C}

\begin{document}

\title{Multiplicative matching of neutral-current deep inelastic scattering processes at next-to-leading order in \pythia~8}

\author{Ilkka Helenius\thanksref{e1,addr1,addr2}\orcidlink{0000-0003-1998-038X}
        \and
        Joni O.~Laulainen\thanksref{e2,addr1,addr2}\orcidlink{0009-0006-9135-6513}
        \and
        Christian T.~Preuss\thanksref{e3,addr3}\orcidlink{0000-0003-1254-0250} 
}

\thankstext{e1}{e-mail: ilkka.m.helenius@jyu.fi}
\thankstext{e2}{e-mail: joni.o.laulainen@jyu.fi (corresponding author)}
\thankstext{e3}{e-mail: preuss@physik.rwth-aachen.de}

\institute{Department of Physics, University of Jyvaskyla, P.O. Box 35, FI-40014 University of Jyvaskyla, Finland\label{addr1}
          \and
          Helsinki Institute of Physics, P.O. Box 64, FI-00014 University of Helsinki, Finland\label{addr2}
          \and
          Institut f{\"u}r Theoretische Teilchenphysik und Kosmologie, RWTH Aachen University, 52056 Aachen, Germany\label{addr3}
}

\date{Received: date / Accepted: date}

\maketitle

\begin{abstract}

We introduce a method for matching the neutral-current deep inelastic scattering process with parton showers in \pythia\ 8, at first order in the strong coupling. The implementation, based on the well-established multiplicative matching approach, is achieved by reweighting leading-order Born-level events and requires that the first parton-shower emission is distributed according to the real matrix-element. 
The method is implemented as an internal matching strategy in the \pythia\ 8 event generator applicable to both currently available parton-shower algorithms, the default one and \vincia. The validity of higher-order corrections is verified with comparisons against existing next-to-leading order simulations. The strategy is used to describe reduced cross-sections measured at the HERA collider, and we find better overall agreement and reduced uncertainties with the matching. 

\end{abstract}
P3H-26-033, MCNET-26-08, TTK-26-13
\vspace{-1em}

\section{Introduction}

Improving the accuracy of Monte Carlo (MC) event generators has been an active field of study, driven by the increasing precision of collider experiments and accomplishments of higher-order calculations in Quantum Chromodynamics (QCD). In particular, the general-purpose event generators \cite{Buckley:2011ms} rely on the perturbative expansion of QCD (pQCD) where the precision of hard-process cross sections can be systematically improved by accounting for higher-order corrections \cite{CTEQ:1993hwr}. The scale evolution of the QCD partons is, in turn, given by the DGLAP evolution equations \cite{Gribov:1972ri, Lipatov:1974qm, Dokshitzer:1977sg, Altarelli:1977zs} which resum collinear divergences up to all orders in $\alphaS$, up to a given logarithmic approximation. In event generators these evolution equations are applied to construct a parton shower (PS) which explicitly generates QCD emissions from initial- and final-state partons taking part in the hard scattering. When combining the PS approach with QCD corrections to the hard-process, potential overlap between the two should be avoided.

Two established approaches are the MC@NLO matching \cite{Frixione:2002ik} and the POWHEG method \cite{Nason:2004rx}. 
In MC@NLO, the corrections are included by subtracting the overlap with the shower explicitly, thereby producing Born-like and real-emission events.
The POWHEG method, on the other hand, achieves NLO accuracy by reweighting Born-level events to the inclusive NLO cross section and generating the first branching according to the real-emission matrix element, akin to matrix-element corrections (MEC) in parton showers. MECs were first adopted for $e^+ e^-$ annihilation in \cite{Bengtsson:1986hr} and later for $W$-production in $pp$-collisions in \cite{Miu:1998ju}, with many more corrections being added to \pythia\ in following years. The two matching strategies have been studied and compared extensively in many generators and collision systems \cite{Hoeche:2011fd,Frixione:2007vw,Alioli:2010xd,Hoche:2010pf,Platzer:2011bc,Alwall:2014hca}.

The application of these methods has been centered on processes studied at the Large Hadron Collider (LHC), but recently applying such algorithms to deep-inelastic scattering (DIS) processes has gained some attention as well, with the main motivation being to provide accurate simulations for future experiments at the Electron-Ion Collider (EIC) \cite{AbdulKhalek:2021gbh}. The POWHEG method for neutral-current (NC) and charged-current DIS \cite{Banfi:2023mhz} has been introduced in the \powhegbox\ framework \cite{Jezo:2015aia,Alioli:2010xd}, with an extension to the polarized case in \cite{Borsa:2024rmh}. Another approach presented in \cite{Jadach:2011cr} introduces a Monte Carlo factorization scheme including NLO corrections. 
Both the additive and multiplicative matching approaches are also available in the Herwig event generator \cite{Bellm:2025pcw}, with the latter applied to DIS processes in \cite{Platzer:2011bc,DErrico:2011wfa}, and more recently to polarized DIS in \cite{Masouminia:2026bhk}. 
A multiplicative matching of the neutral-current DIS process with a formally next-to-leading-logarithmic accurate parton-shower algorithm \cite{vanBeekveld:2023chs} has recently been established in the context of the PanScales framework \cite{vanBeekveld:2025lpz}, adapting the projection-to-Born approach \cite{Cacciari:2015jma}. Prior to adaptation to shower matching approaches, this method has been used in the context of DIS fixed-order calculations, for example N$^3$LO jet production \cite{Currie:2018fgr,Gehrmann:2018odt} in the \textsc{NNLOjet} program \cite{NNLOJET:2025rno}, as well as NNLO jet production in the \textsc{disorder} \cite{Karlberg:2024hnl} and \textsc{poldis} \cite{Borsa:2022cap} programs. 
In the context of the \sherpa\ framework \cite{Sherpa:2024mfk}, the MC@NLO and projection-to-Born methods provided the basis for simulations of the DIS process at next-to-next-to-leading order (NNLO) accuracy in \cite{Hoche:2018gti} and for a multi-jet merged calculation with up to 3 jets at NLO in \cite{Meinzinger:2025pam}. 
In early versions, \sherpa\ also implemented the POWHEG matching scheme, including DIS one-jet production \cite{Hoche:2010pf}.

In \pythia\ 8 \cite{Bierlich:2022pfr}, DIS is available via internal processes at LO with PS \cite{Cabouat:2017rzi}, which can be combined with higher-multiplicity tree-level matrix elements via a dedicated merging scheme \cite{Helenius:2024wjg}.
For NLO simulations, \pythia\ 8 has so far relied on external input.
A noteworthy exception to this is the production of dijet events in color-singlet decays, for which a multiplicative matching to the NLO width has been introduced about 40 years ago \cite{Bengtsson:1986et}, marking an early example of the multiplicative matching strategy nowadays known as POWHEG. 

Dedicated matrix-element generators capable of producing process-level events at higher orders in $\alphaS$, such as \textsc{MadGraph5\_aMC@NLO} \cite{Alwall:2014hca} or \powhegbox\ \cite{Alioli:2010xd}, need interfacing with a general-purpose generator to generate parton showers and hadronize the event. A popular option for this is \pythia, where native NLO matching capabilities have been sparse. Here, we present a multiplicative matching implementation for the charged-lepton neutral-current DIS process. This marks the first fully-internal NLO simulation in \pythia\ for a process with hadrons in the initial state.
To this end, we adopt the projection-to-Born approach for the multiplicative NLO factor. 
The standard NLO hard-coefficient functions are acquired analytically and, based on the real-emission amplitudes part of this standard calculation, the required matrix-element corrections to the default ``simple shower'' and \vincia\ shower algorithms are derived.

Specifically, our implementation is based on the following assumptions:
\begin{itemize}
\item no shower recoil is absorbed by the scattered electron, so that the Born kinematics are retained fully differentially throughout the shower;
\item the Born-local inclusive NLO $K$-factor can be calculated analytically through NLO structure functions up to numerical convolutions with the PDFs;
\item matrix-element corrections are performed for the photon-induced scattering processes only; and
\item NLO accuracy is achieved only for flavor-agnostic observables.
\end{itemize}

Our proof-of-concept implementation establishes a new framework for higher-order calculations in \pythia, which may be extended to other processes and parton-shower algorithms, such as the \textsc{Apollo} shower \cite{Preuss:2024vyu}, in the future.
Moreover, the framework is general enough so that any of the above assumptions may be relaxed in the future.
Nevertheless, we here focus on establishing the framework and adopting it for a simple process.
We validate the matching scheme with comparisons against \sherpa\ and \powhegbox\ frameworks and showcase predictions of reduced cross-sections of inclusive NC DIS process measured by HERA, highlighting improved agreement to data and reduced scale uncertainties compared to the LO case. 
The developments discussed here have been made public in the \pythia\ 8.317 release, with some technical fixes in the version 8.318.

To establish the conventions used throughout this work, we start by reviewing the relevant theoretical foundations in section \ref{sec:theory}. To this end, we reproduce the calculation of structure functions at NLO, identify the real-emissions amplitudes needed for the MECs, introduce the \pythia\ and \vincia\ shower algorithms, and outline the multiplicative matching strategy employed here. 
Section \ref{sec:implementation} describes how the strategy is implemented in \pythia, with parton-shower matrix-element corrections as the main focus. In section \ref{sec:validation}, the implementation's ability to describe inclusive observables at NLO is investigated and compared to existing calculations in \sherpa\ and \powhegbox. The impact of NLO corrections on the description of reduced cross-sections measured at HERA is demonstrated in section \ref{sec:results} together with predictions for leading-jet kinematics at the EIC.

\section{Theoretical background}
\label{sec:theory}

Parton densities and DIS jet cross-sections have been extensively studied beyond leading order \cite{Altarelli:1979ub,Furmanski:1981cw,Gluck:1994uf,Moch:1999eb,Daleo:2006xa} which we follow in the derivation of the NLO structure functions. The matching strategy considered in this work requires the computation of the NLO weight fully differentially in the Born-level phase space. This can be achieved by considering the DIS cross section in terms of the structure functions at NLO. To establish an explicit connection between the standard NLO structure functions and matrix-element correction factors in the parton shower, we summarize the calculation of the relevant matrix elements from the hadronic tensor below, in Section \ref{sec:NLODIS}.
The relevant structure functions are constructed in collinear factorization as convolutions of PDFs and coefficient functions, including the neutral-current process. 
We present the calculation in dimensional regularization with $D=4-2\varepsilon$ dimensions and use the $\MSBAR$ scheme, in which the remaining collinear divergences and certain finite contributions are absorbed into the definition of PDFs.
Subsequently, we summarize the parton-shower algorithms available in \pythia\ 8 in section \ref{sec:showers}, before we continue to describe the philosophy of well-established multiplicative matching schemes in section \ref{sec:matching}.

\subsection{Inclusive DIS at NLO}
\label{sec:NLODIS}
\begin{figure}
    \centering
    \includegraphics[width=0.75\linewidth]{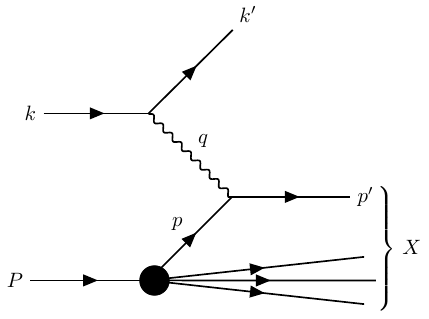}
    \caption{The inclusive DIS process $ep\to eX$ and momentum assignments. }
    \label{fig:DIS_momenta2}
\end{figure}

The neutral-current electron-proton DIS cross section can be written in terms of the structure functions $F_1$, $F_L = F_2 - 2xF_1$ and $F_3$ as \cite{ParticleDataGroup:2024cfk}
\begin{equation}
\begin{split}
    \frac{\D^2 \sigma}{\D x \D Q^2} &= \frac{4\pi \alpha^2}{xQ^4} \Bigg\{ \left(1-y+\frac{y^2}{2}\right)F_2(x,\MUF^2) \\ &- \frac{y^2}{2}F_L(x,\MUF^2) \mp \left(y-\frac{y^2}{2}\right)xF_3(x,\MUF^2) \Bigg\}\,,
\end{split}
\label{eq:dsigma_dxdQ2}
\end{equation}
where the Lorentz-invariant kinematic variables are $x = Q^2/2P\cdot q $, inelasticity $y = P\cdot q / P\cdot k$ and virtuality $Q^2 = -q^2 = -(k-k')^2$, with $P$, $k$, $k'$ and $q$ being the 4-momenta of proton, incoming electron, outgoing electron and photon, respectively, as depicted in Fig.~\ref{fig:DIS_momenta2}. The invariant mass of the hadronic final state $X$ is given by $W^2 = (P+q)^2$, $\MUF$ is the factorization scale, and $\alpha$ is the QED coupling constant. The sign of the third term is $-$ for an initial-state $e^+$ and $+$ for $e^-$. 

The differential cross section can be written as a contraction of the leptonic tensor $L^{\mu\nu}$ and the hadronic tensor $W_{\mu\nu}$, 
\begin{equation}
    \frac{\D^2 \sigma}{\D x \D Q^2} = \frac{2\pi y\alpha^2}{sxQ^4} L^{\mu\nu} W_{\mu\nu} \,.
\label{eq:hadrTensor}
\end{equation}
In the collinear factorization framework \cite{Collins:1989gx}, this can be written as a convolution over the partonic cross section and the parton distribution functions $f_i$, 
\begin{align}
    \label{eq:LW_factorized}
    \frac{\D^2 \sigma}{\D x \D Q^2} &= \sum_i f_i \otimes \frac{\D^2 \hat{\sigma}^i}{\D x \D Q^2} =\sum_i \int_0^1 \frac{\D \xi}{\xi} f_i(\xi) \frac{\D^2 \hat{\sigma}^i}{\D x \D Q^2}
\end{align}
where the sum runs over partons $i=q,g$ and $\xi$ is the fraction of proton momentum carried by the parton participating in the hard scattering. The hadronic tensor can be written in terms of dimensionless structure functions as in Ref.~\cite{Furmanski:1981cw}
\begin{equation}
\begin{split}    
    MW_{\mu\nu} = &- \left(g_{\mu\nu} - \frac{q_\mu q_\nu}{q^2}\right)F_1(x, \MUF^2) \\ 
    &+ \left(P_{\mu} - \frac{P\cdot q}{q^2}q_\mu\right) \left( P_\nu - \frac{P\cdot q}{q^2}q_\nu\right)\frac{F_2(x, \MUF^2)}{P\cdot q} \\
    &- i\epsilon_{\mu\nu\alpha\beta} \frac{P^\alpha q^\beta}{2P\cdot q}F_3(x, \MUF^2) \,,
\end{split}
\label{eq:hadronic_tensor}
\end{equation}
where the inclusion of weak interactions through $Z$-boson exchange and interference with photon exchange give rise to an antisymmetric part in the hadronic tensor of Eq.~\eqref{eq:hadronic_tensor}, resulting in additional contributions to the structure functions $F_1$ and $F_2$ and the parity-violating structure function $F_3$. 

\begin{figure}[t]
    \centering
    \begin{subfigure}{0.15\textwidth}
        \centering
        \includegraphics[width=\linewidth]{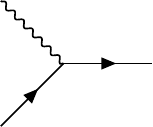}
        \caption{$A^{q(B)}$}
    \end{subfigure}
    \hspace{2em}
    \begin{subfigure}{0.15\textwidth}
        \centering
        \includegraphics[width=\linewidth]{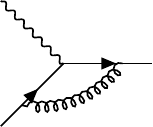}
        \caption{$A^{q(V)}$}
    \end{subfigure}
    
    \vspace{0.5em}
    
    \begin{subfigure}{0.15\textwidth}
        \centering
        \includegraphics[width=\linewidth]{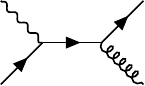}
        \caption{$A^{q(R)}_s$}
    \end{subfigure}
    \hspace{2em}
    \begin{subfigure}{0.15\textwidth}
        \centering
        \includegraphics[width=\linewidth]{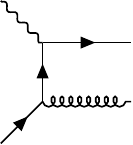}
        \caption{$A^{q(R)}_t$}
    \end{subfigure}
    
    \vspace{0.5em}
    
    \begin{subfigure}{0.15\textwidth}
        \centering
        \includegraphics[width=\linewidth]{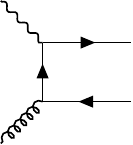}
        \caption{$A^{g(R)}_t$}
    \end{subfigure}
    \hspace{2em}
    \begin{subfigure}{0.15\textwidth}
        \centering
        \includegraphics[width=\linewidth]{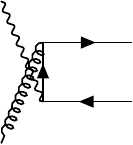}
        \caption{$A^{g(R)}_u$}
    \end{subfigure}
    
    \caption{The non-zero diagrams and the corresponding amplitudes contributing to deep inelastic scattering at next-to-leading order. }
    \label{fig:DIS_NLO}
\end{figure}

Projecting the structure functions in terms of contractions with the partonic tensor (in $D=4-2\varepsilon$ dimensions) gives
\begin{align}
\begin{split}
    \label{eq:F2_contractions}
    \frac{F_2(x,\MUF^2)}{x} &= \frac{1}{1-\varepsilon}\bigg[-g^{\mu\nu}(MW_{\mu\nu}) \\ 
    &+ (3-2\varepsilon)\frac{4x^2}{Q^2}P^\mu P^\nu(MW_{\mu\nu}) \bigg]
\end{split}\\
\begin{split}
    \label{eq:FL_contractions}
    \frac{F_L(x,\MUF^2)}{2x} &= \frac{4x^2}{Q^2}P^\mu P^\nu(MW_{\mu\nu})
\end{split}\\
\begin{split}
    \label{eq:F3_contractions}
    F_3(x,\MUF^2) &= \frac{2x}{Q^2}i\epsilon^{\mu\nu\alpha\beta}P_\alpha q_\beta (MW_{\mu\nu}).
\end{split}
\end{align}
Extracting the NLO structure functions requires the calculation of these contractions with the diagrams contributing to the unpolarized partonic NC DIS process, shown in Fig.~\ref{fig:DIS_NLO}. 
Squaring the amplitudes and summing and averaging over final- and initial-state colors, spins, and polarizations gives the following non-zero contractions for the electromagnetic interaction: the Born-level amplitude
\begin{equation}
    g^{\mu\nu}\overline{A^{q(B)}_\mu A^{q(B)*}_\nu} = -2 e_q^2 Q^2 (1-\varepsilon)
\label{eq:bornME}
\end{equation}
and the virtual part 
\begin{equation}
\begin{split}
\label{eq:virtual}
    g^{\mu\nu}2\mathrm{Re} \big( \overline{A^{q(V)}_{\mu}A^{q(B)*}_\nu} \big) &= -e_q^2 \frac{\alphaS}{4\pi} \bigg( \frac{4\pi\mu^2}{Q^2}\bigg)^\varepsilon \frac{4Q^2(1-\varepsilon)}{\Gamma (1-\varepsilon)} \\
    &\times \CF \bigg[ -\frac{2}{\varepsilon^2} - \frac{3}{\varepsilon} - 8 + \mathcal{O}(\varepsilon) \bigg].
\end{split}
\end{equation}
The real-emission contributions are defined in terms of invariants $\hat{s} = (p+q)^2$, $\hat{t} = (p-p_2)^2$ and $\hat{u} = (p-p_1)^2$, where $p_1$ and $p_2$ are the momenta of the outgoing partons, as depicted in the real-emission graphs (c) through (f) of Fig.\ref{fig:DIS_NLO}. For the quark-initiated contributions we have
\begin{align}
\begin{split}
\label{eq:real-q-g}
    g^{\mu\nu}\overline{A^{q(R)}_\mu A^{q(R)*}_\nu} &= -\frac{e_q^2}{2} 4\pi\alphaS (\mu^2)^\varepsilon (1-\varepsilon) \\
    &\times \CF 8\bigg[ \left( -\frac{\hat{s}}{\hat{t}} -\frac{\hat{t}}{\hat{s}} \right) (1-\varepsilon) \\
    &\qquad\quad + \frac{2Q^2\hat{u}}{\hat{s}\hat{t}} + 2\varepsilon + \mathcal{O}(\varepsilon^2) \bigg] \,,
\end{split}\\
\label{eq:real-q-pp}
    p^\mu p^\nu \overline{A^{q(R)}_\mu A^{q(R)*}_\nu} &= \frac{e_q^2}{2} 4\pi\alphaS (\mu^2)^\varepsilon (1-\varepsilon) \TR 8 \bigg[ -\frac{\hat{u}}{2} \bigg] \,,
\end{align}
and for the gluon contributions to real emissions
\begin{align}
\begin{split}
\label{eq:real-g-g}
    g^{\mu\nu}\overline{A^{g(R)}_\mu A^{g(R)*}_\nu} &= -\frac{e_q^2}{2} 4\pi\alphaS (\mu^2)^\varepsilon (1-\varepsilon) \\
    &\times \CF 8\bigg[ \left( \frac{\hat{t}}{\hat{u}} +\frac{\hat{u}}{\hat{t}} \right) (1-\varepsilon) \\
    &\qquad\quad - \frac{2Q^2\hat{s}}{\hat{t}\hat{u}} + \mathcal{O}(\varepsilon^2) \bigg] \,,
\end{split}\\
\label{eq:real-g-pp}
    p^\mu p^\nu \overline{A^{g(R)}_\mu A^{g(R)*}_\nu} &= \frac{e_q^2}{2} 4\pi\alphaS (\mu^2)^\varepsilon (1-\varepsilon) \TR 8 \bigg[ \hat{s} \bigg] \,,
\end{align}
where the color factors are $\CF = 4/3$ and $\TR = 1/2$. Similar contractions are obtained for $Z$-interaction and interference contributions. 

Combining these with the appropriate $2\to1$ and $2\to2$ phase-space elements provides the required contractions of the hadronic tensor from which we can construct the neutral-current structure functions to the order $\alphaS$. Notice that the explicit poles in virtual correction in Eq.~\ref{eq:virtual} eventually cancel out with real corrections after performing the phase-space integration and the remaining collinear singularity is absorbed into definition of the PDFs. Including the weak-interaction contributions gives $F_2^\mathrm{NC}$ as
\begin{align}
    \nonumber
    \frac{F_2^\mathrm{NC}(x, \MUF^2)}{x} &= \sum_{q} \Big[e_f^2 e_q^2 + 2\eta e_f e_q v_f v_q \\
    \nonumber
    &\qquad  + \eta^2(v_f^2 + a_f^2)(v_q^2 + a_q^2)\Big] \\
    \label{eq:F2NC}
    &\times \Big\{ \Big(q(x, \MUF^2)+\bar{q}(x, \MUF^2)\Big)\\
    \nonumber
    &\qquad \otimes \Big[ 1 + \frac{\alphaS}{2\pi} \Big(  C^{\MSBAR}_{2,q}(z) + \hat{P}_{qq}(z) \log \Big(\frac{Q^2}{\MUF^2}\Big) \Big) \Big] \\
    \nonumber
    &\quad + 2 g(x, \MUF^2) \\
    \nonumber
    &\qquad \otimes \frac{\alphaS}{2\pi} \Big[ C^{\MSBAR}_{2,g}(z) + \hat{P}_{qg}(z) \log \Big(\frac{Q^2}{\MUF^2}\Big) \Big] \Big\},
\end{align}
where 
\begin{equation}
\eta = (16\sin^2 \theta_W \cos^2 \theta_W)^{-1} Q^2 / (Q^2 + M_Z^2)
\end{equation}
contains the ratio of the coupling to the corresponding propagator through the weak mixing angle $\theta_W$. At leading order, the structure functions are sensitive only to the electromagnetic coupling $e = \sqrt{4\pi\alpha}$ and, through $Z$-exchange, the electroweak vector coupling $v_f = a_f - 4e_f \sin \theta_W$ and axial coupling $a_f = \pm 1$ for fermion $f$. 
The $\alphaS$-corrections are proportional to the coefficient functions in the $\MSBAR$-scheme, 
\begin{align}
\begin{split}
    C_{2,q}^{\MSBAR}(z) &= \CF \Bigg\{ (1+z^2) \left( \frac{\log (1-z)}{1-z} \right)_+ - \frac{3}{2} \frac{1}{(1-z)_+} \\
    &\qquad - \frac{1+z^2}{1-z} \log z + 3 + 2z \\
    &\qquad - \Big( \frac{9}{2} + \frac{\pi^2}{3} \Big) \delta(1-z) \Bigg\}\,,
\end{split}\\
\begin{split}
    C_{2,g}^{\MSBAR}(z) &= \TR \Bigg\{ (z^2 + (1-z)^2) \log \Big( \frac{1-z}{z} \Big) \\
    &\qquad + 8z(1-z) - 1 \Bigg\} \,,
\end{split}
\end{align}
where $z = \frac{Q^2}{2p\cdot q} = \frac{x}{\xi}$. These hard coefficients are standard results of pQCD, and are equal to the expressions given in, e.g., Appendix I of Ref. \cite{Furmanski:1981cw} and Eqs. (8) and (15) of Ref. \cite{Gluck:1994uf}. 
The factorization-scale dependence is explicit in terms containing the regularized splitting functions
\begin{align}
    \hat{P}_{qq}(z) &= \CF \Bigg[ \left( \frac{1+z^2}{1-z} \right)_+ + \frac{3}{2}\delta(1-z) \Bigg], \\
    \hat{P}_{qg}(z) &= \TR \left(z^2 + (1-z)^2\right).
\end{align}
Similarly, for the structure function $F_L$ we get 
\begin{equation}
\begin{split}
    \frac{F_L^\mathrm{NC}(x, \MUF^2)}{x} &= \sum_{q} \Big[e_f^2 e_q^2 + 2\eta e_f e_q v_f v_q \\
    &\qquad  + \eta^2(v_f^2 + a_f^2)(v_q^2 + a_q^2)\Big] \\
    &\times \Big\{ \Big(q(x, \MUF^2)+\bar{q}(x, \MUF^2)\Big) \otimes \frac{\alphaS}{2\pi} \Big[  C^{\MSBAR}_{L,q}(z) \Big] \\
    &\quad + 2 g(x, \MUF^2) \otimes \frac{\alphaS}{2\pi} \Big[ C^{\MSBAR}_{L,g}(z) \Big] \Big\},
\end{split}
\label{eq:FLNC}
\end{equation}
with 
\begin{align}
    C_{L,q}^{\MSBAR}(z) &= \CF 2z, \\
    C_{L,g}^{\MSBAR}(z) &= \TR 4z(1-z).
\end{align}
Finally, for the structure function $F_3$ we find
\begin{align}
    \nonumber
    F_3^\mathrm{NC}(x, \MUF^2) &= \sum_{q} \left[\eta 2e_f v_f e_q v_q + \eta^2 4 v_f a_f v_q a_q \right] \\
    \label{eq:F3NC}
    &\times \Big\{ \Big(q(x, \MUF^2)-\bar{q}(x, \MUF^2)\Big) \\
    \nonumber
    &\qquad \otimes \Big[1 + \frac{\alphaS}{2\pi} \Big( C^{\MSBAR}_{3,q}(z) + \hat{P}_{qg}(z) \log \Big(\frac{Q^2}{\MUF^2}\Big) \Big) \Big] \Big\}, 
\end{align}
with 
\begin{equation}
    C_{3,q}^{\MSBAR}(z) = C_{2,q}^{\MSBAR}(z) - \CF (1+z).
\end{equation}
Combining these gives us finally the full inclusive DIS cross section to the order $\alphaS$ that we will use to derive the NLO correction.

In addition to the inclusive NLO cross section above, we will also need the explicit matrix elements for quark- and gluon-initiated real-emission diagrams with an incoming virtual photon. As pointed out in Ref.~\cite{Catani:1996vz}, these can be obtained from a particular combination of contractions with squared amplitudes in Eqs.~(\ref{eq:real-q-g}) -- (\ref{eq:real-g-pp}).

The squared matrix element for the quark-induced channel is
\begin{equation}
\begin{split}
    \vert \mathcal{M}_{q\gamma^*\to qg}^{(0)}\vert^2 &= \vert \mathcal{M}_{q\gamma^*\to q}^{(0)}\vert^2 \frac{8\pi\alphaS}{Q^2} (\mu^2)^\varepsilon \CF \\ 
    &\times \bigg[\left( -\frac{\hat{s}}{\hat{t}} -\frac{\hat{t}}{\hat{s}} \right) (1-\varepsilon) + \frac{2Q^2\hat{u}}{\hat{s}\hat{t}} \\
    &\qquad - \frac{6Q^2\hat{u}}{(\hat{s}+Q^2)^2} + 2\varepsilon + \mathcal{O}(\varepsilon^2) \bigg]\,
\end{split}
\label{eq:ME_Rq}
\end{equation}
and the gluon-induced channel
\begin{equation}
\begin{split}
    \vert \mathcal{M}_{g\gamma^*\to q\bar{q}}^{(0)}\vert^2 &= \vert \mathcal{M}_{q\gamma^*\to q}^{(0)}\vert^2 \frac{8\pi\alphaS}{Q^2} (\mu^2)^\varepsilon \TR \\
    &\times \bigg[ \bigg( \frac{\hat{u}}{\hat{t}} +\frac{\hat{t}}{\hat{u}} \bigg) \bigg(1-\varepsilon - \frac{2Q^2\hat{s}}{(\hat{s}+Q^2)^2} \bigg)\\
    &\qquad + \frac{8Q^2\hat{s}}{(\hat{s}+Q^2)^2} + \mathcal{O}(\varepsilon^2) \Bigg]\,,
\end{split}
\label{eq:ME_Rg}
\end{equation}
for which the parts corresponding to contractions with $g^{\mu\nu}$ can be found in e.g. \cite{Altarelli:1979ub}. 
Here, the leading-order Born-level matrix element is
\begin{equation}
    \vert \mathcal{M}_{q\gamma^*\to q}^{(0)}\vert^2 = -2e_q^2 Q^2 (1-\epsilon) \,. 
\end{equation}
These expressions will be used in section \ref{sec:implementation} to construct the matrix-element corrections required in the matching strategy.

\subsection{Showers in \pythia\ 8}
\label{sec:showers}
Parton showers describe the transition from a low-multiplicity hard-scattering event to a low-energy state with many partons, ordered in some resolution variable $t$.
By construction, a parton-shower algorithm is typically manifestly unitary, i.e., it produces additional partons from the Born-level state by applying consecutive branchings through a unitary operation and thus does not modify the inclusive Born-level cross section.
This can be cast into a parton-shower operator $\mathcal{F}$, defined recursively as
\begin{equation}
\begin{split}
    &\mathcal{F}_{n}(O,\Phi_n) = \Delta_n(t_n,t_\mathrm{c})\, O(\Phi_n) \\
    &+ \int\limits^{t_n}_{t_\mathrm{c}}\, K_{n\to n+1}(\Phi_{n+1})\, \Delta_n(t_n,t)\, \mathcal{F}_{n+1}(O,\Phi_{n+1})\, \D\Phi_{+1} \,,
\end{split}
\label{eq:PS_operator}
\end{equation}
where $O$ denotes an arbitrary, infrared- and collinear-safe observable.
Here, the quantities $K_{n\to n+1}$, $\Phi_{n}$, and $\Phi_{+1}$ denote the shower branching kernel, $n$-particle phase space, and one-particle branching phase space, respectively.
The shower operator $\mathcal{F}$ contains the no-branching probability,
\begin{equation}
    \Delta_n(t_n,t_\mathrm{c}) = \exp\left\{-\int\limits^{t_n}_{t_\mathrm{c}}\, K_{n\to n+1}\, \D\Phi_{+1} \right\}\, ,
\end{equation}
which encodes the probability for no branching to occur between two evolution scales $t_n$ and $t_c$.

The explicit definition of the shower evolution variable $t$, the form of the branching kernels $K$, and the phase-space factorization underlying $\D\Phi_{+1}$ depend on the parton-shower algorithm at hand.
In its present release, \pythia\ 8 currently provides\footnote{Note that the Dire parton-shower algorithm has been removed in \pythia\ 8.316. } two independent parton showers, namely the default transverse-momentum ordered ``simple shower'' \cite{Sjostrand:2004ef} and the \vincia\ antenna shower \cite{Brooks:2020upa}, which we will briefly review below.

Both showers implement a local antenna recoil scheme for initial-final color dipoles, but differ slightly in the treatment of final-final dipoles, where the simple shower implements a local dipole recoil scheme, whereas \vincia\ implements emitter-recoiler agnostic antenna kinematics. In the context of continuous global observables, both showers formally resum leading-logarithmic contributions, as they employ a local transverse recoil and are ordered in a notion of transverse momentum \cite{vanBeekveld:2023chs}. An extension to parton showers with higher logarithmic accuracy for this class of observables, such as Apollo \cite{Preuss:2024vyu}, is left for future work. 

\paragraph{The simple shower} in \pythia\ \cite{Sjostrand:2004ef} is ordered in transverse momentum defined in terms of the momentum fraction $z$ and the virtuality of the appropriate emitting parton $\pm m_{a/b}^2$ in a $a \to bc$ splitting, 
\begin{equation}
    t = \begin{cases}
        z(1-z)m_{a}^2 & \text{FSR} \\ (1-z)(-m_{b}^2) & \text{ISR}
    \end{cases}.
\end{equation}
Branching kernels in the simple shower are defined in terms of the standard (unregularized) DGLAP splitting kernels,
\begin{align}
    K_{a\to bc}(t,z,\phi) &= \frac{\alphaS(t)}{2\pi} \frac{P_{a\to bc}(z)}{t}\,,
\end{align}
given by, 
\begin{align}
    P_{q\to qg}(z) &= \CF \frac{1+z^2}{1-z} \,, \\
    P_{g\to gg}(z) &= \CA \frac{(1-z(1-z))^2}{z(1-z)} \,,\\
    P_{g\to q\bar{q}}(z) &= \TR (z^2 + (1-z)^2) \,,
\end{align}
where $\CA = 3$. By default, FSR branchings in the simple shower recoil locally in the dipole, with the transverse recoil generated by the emission being absorbed by the emitter and a longitudinally recoiling spectator. 
ISR branchings, on the other hand, would by default recoil against the entire final state. 
This procedure is, however, in conflict with DIS kinematics. An alternative evolution scheme, known as the ``dipole-recoil'' scheme, exists in which the transverse recoil in ISR branchings is taken only by the color-connected final-state parton \cite{Cabouat:2017rzi}.
This procedure guarantees to leave the DIS variables derived from the outgoing lepton four-momentum invariant upon generation of all PS branchings. 

\paragraph{The \textsc{Vincia} antenna shower} \cite{Brooks:2020upa} implements an evolution in a generalization of the Ariadne transverse-momentum definition \cite{Lonnblad:1992tz},
\begin{equation}
    t = \frac{\bar{q}_{bc}^2\bar{q}_{cd}^2}{s_\mathrm{max}}\,, \quad \bar{q}_{bc}^2 = \begin{cases}
        (p_b+p_c)^2-m_a^2 & a~\text{final}\\
        -(p_b-p_c)^2+m_a^2 & a~\text{initial}
    \end{cases}
\end{equation}
where $s_\mathrm{max}$ denotes the maximal invariant mass of the antenna and $m_a^2$ is the invariant mass of the radiating parton $a$.
In \textsc{Vincia}, branching kernels are replaced by antenna functions, given in terms of dimensionless invariants $y_{bc}$ and $y_{cd}$,
\begin{equation}
    K_{a\to bc}(t,z,\phi) = \frac{\alphaS(t)}{2\uppi} C_{bc} A_{c/ad}(y_{bc},y_{cd})\,,
\end{equation}
with $d$ denoting the color-connected parton in the antenna $a$-$d$ and $C_{bc}$ the corresponding color factor of the splitting. 
For massless partons, the final-final antenna functions read
\begin{align}
    A_{g/q\bar{q}}(y_{bc}, y_{cd}) &= \frac{1}{s_{bcd}}\frac{(1-y_{bc})^2+(1-y_{cd})^2}{y_{bc}y_{cd}} + \ldots\,,\\
    A_{g/qg}(y_{bc}, y_{cd}) &= \frac{1}{s_{bcd}}\frac{(1-y_{bc})^3+(1-y_{cd})^2}{y_{bc}y_{cd}} + \ldots\,,\\
    A_{g/gg}(y_{bc}, y_{cd}) &= \frac{1}{s_{bcd}}\frac{(1-y_{bc})^3+(1-y_{cd})^3}{y_{bc}y_{cd}} + \ldots\,,\\
    A_{q/gX}(y_{bc}, y_{cd}) &= \frac{1}{s_{bcd}}\frac{y_{bd}^2+y_{cd}^2}{2y_{bc}} + \ldots\,,
\end{align}
where non-singular terms have been omitted. 
For initial-final and initial-initial antennae, appropriate crossings of the antenna functions above are used.
The full set of antenna functions implemented in \textsc{Vincia} can be found in the appendix of \cite{Brooks:2020upa}.

\paragraph{The starting scale} of the shower evolution should be set to the factorization scale of the hard process, $\MUF$, motivated by factorization arguments.
This approach misses phase-space regions with $t > \MUF^2$, which the shower may in principle be able to describe well, especially when matrix-element corrections are employed.
The situation is especially restricting in the case of DIS, where the factorization scale is often chosen proportional to the negative photon momentum transfer, $Q^2$, which typically acquires small values of the order of a few GeV$^2$. 
The recommended approach in \pythia\ for many processes is therefore to let the shower start at the phase-space maximum, given by the hadronic center-of-mass energy squared, $s$, \cite{Plehn:2005cq} including a dampening factor for high-virtuality emissions in cases where MECs are not available.
While these ``power showers'' formally violate factorization theorems, phenomenologically, they often yield a reasonable description of experimental data. 
In particular, this choice does not affect the NLO accuracy of the method for inclusive observables, as the parton shower remains unitary regardless of the starting scale and all PDFs in the Born-local NLO weight are evaluated at the factorization scale. 
In the following, we will therefore limit the discussion to power showers to leave the focus on the calculation and implementation of the NLO correction.
We wish to emphasize, however, that a restriction of the shower evolution to the region $t < \MUF^2$ may straightforwardly be achieved through multi-jet merging techniques, for which \pythia\ provides a dedicated implementation in DIS processes \cite{Helenius:2024wjg}.
We leave the combination of NLO matching and multi-jet merging in DIS processes with \pythia\ to future work.

\subsection{Multiplicative matching}
\label{sec:matching}

In this section, we will briefly review the schematic
derivation of a multiplicative matching strategy, following the arguments in \cite{Bengtsson:1986hr,Nason:2004rx,Frixione:2007vw,Hoche:2010pf,Platzer:2011bc}. We start from the expectation value of an infrared-safe observable $O$ at NLO, and the first-order expansion of the parton-shower prediction for the observable. 

On the one hand, the expectation value of $O$ can be written at NLO schematically as the projection-to-Born prediction \cite{Cacciari:2015jma}
\begin{equation}
\begin{split}
    \langle O \rangle_\mathrm{NLO} &= \int \bar{B}(\Phi_B) \, O(\Phi_B)\, \D\Phi_B \\
    &+ \int R(\Phi_R)\, \left[O(\Phi_R) - O(\Phi_B)\right]\, \D\Phi_R \,,
\end{split}
\label{eq:O_NLO}
\end{equation}
where the function $\bar{B}(\Phi_B)$ encodes the inclusive NLO rate, differential in the Born momenta $\Phi_B$,
\begin{equation}
    \bar{B}(\Phi_B) = B(\Phi_B) + V(\Phi_B) + \int R(\Phi_R)\, \D\Phi_{+1} \,.
\label{eq:BBar}
\end{equation}
Here, the Born-level contribution $B(\Phi_B)$ is the leading-order tree-level amplitude squared, and $V(\Phi_B)$ the virtual correction, both evaluated at a Born-level phase-space point $\Phi_B$. 
The quantity $R(\Phi_R)$ is the real-emission contribution evaluated at a phase-space point of a real-emission $\Phi_R$, and the real-emission phase-space differential has been factorized as $\D\Phi_R = \D\Phi_B \D\Phi_{+1}$. 
A suitable regularisation of the implicit singularities in $R$ and explicit singularities in $V$ is assumed in Eq.~\eqref{eq:BBar}.

On the other hand, the parton-shower prediction for the observable $O$ can be written with the definition of the parton-shower operator in Eq.~\eqref{eq:PS_operator} as
\begin{equation}
    \label{eq:O_PS}
    \langle O\rangle_\mathrm{PS} = \int B(\Phi_B)\, \mathcal{F}(O,\Phi_B)\, \D\Phi_B \,.
\end{equation}
Considering only the first emission, we have
\begin{equation}
\begin{split}
    \mathcal{F}_0(O, \Phi_B) &= \Delta_0(t_0,t_\mathrm{c})\, O(\Phi_B) \\
    &+ \int\limits^{t_0}_{t_\mathrm{c}}\, K_{B\to R}(\Phi_{R})\, \Delta_0(t_0,t)\, \mathcal{F}_1(O, \Phi_{R})\, \D\Phi_{+1} \,,
\end{split}
\end{equation}
where the subscripts $B$ and $R$ refer to the $n=0$ Born and $n=1$ real-emission states, respectively. To match the prediction at NLO accuracy, the expression is expanded as a series in powers of $\alphaS$. 
The $\mathcal{O}(\alphaS)$ contribution of the first emission is then
\begin{equation}
\begin{split}
    \mathcal{F}_0^{(1)}(O, \Phi_B) &= O(\Phi_B) \\
    &+ \int\limits^{t_0}_{t_\mathrm{c}}\, K_{B\to R}(\Phi_{R})\, \left[O(\Phi_R) - O(\Phi_B)\right]\, \mathrlap{{} \D\Phi_{+1} \,.}
\end{split}
\label{eq:F1_NLO}
\end{equation}
Substituting the first-emission contribution of Eq.~\eqref{eq:F1_NLO} to the parton-shower prediction of \eqref{eq:O_NLO} and collecting terms proportional only to the first power of $\alphaS$, we get
\begin{equation}
\begin{split}
    \langle O\rangle_\mathrm{PS}^{(1)} &= \int \, B(\Phi_B) O(\Phi_B)\, \D\Phi_B\, \\
    &+ \int\, B(\Phi_B) K_{B\to R}(\Phi_{R})\, \left[O(\Phi_R) - O(\Phi_B)\right]\, \D\Phi_R \\
    &- \int\limits_{0}^{t_c}\, B(\Phi_B) K_{B\to R}(\Phi_{R})\, \left[O(\Phi_R) - O(\Phi_B)\right]\, \D\Phi_R \,,
\end{split}
\end{equation}
where the last term is the residual contribution arising from the shower cutoff.
It is a finite power correction in $t_c$ and vanishes in the limit $t_c \to 0$. 
For sufficiently small cutoffs, the remainder is negligible and hence omitted in further considerations.

We can now match this parton-shower prediction with the NLO prediction of Eq.~\eqref{eq:O_NLO} via substituting the Born factor $B(\Phi_{B})$ by the NLO factor $\bar{B}(\Phi_{B})$ in the first term, and substituting the shower kernel $K_{B\to R}(\Phi_R)$ by the ratio $R(\Phi_{R}) / B(\Phi_{B})$ in the second term. In practice, the first substitution is achieved by a simple reweighting, whereas the second substitution requires utilizing matrix-element corrections to the shower kernels. 
The technical implementation of the latter corrections is discussed in detail in section \ref{sec:showers}. 
With these substitutions, we arrive at the first-order expansion of the multiplicatively matched NLOPS prediction,
\begin{align}
\begin{split}
    &\langle O\rangle_\mathrm{NLOPS}^{(1)} = \int \, \bar{B}(\Phi_B) O(\Phi_B)\, \D\Phi_B\, \\
    \label{eq:NLOPS}
    &\qquad + \int\limits_{t_c}^t\, B(\Phi_B) K_{B\to R}(\Phi_{R})\, \left[O(\Phi_R) - O(\Phi_B)\right]\, \D\Phi_R\,,
\end{split}
\end{align}
which matches the projection-to-Born NLO prediction in Eq.~\eqref{eq:O_NLO} up to power corrections in $t_c$, discussed above.

The NLO matched prediction
\begin{equation}
    \langle O\rangle_\mathrm{PS} = \int\, \bar{B}(\Phi_B)\, \mathcal{F}^\mathrm{MEC}(O,\Phi_B)\, \D \Phi_B
\end{equation}
ensures that any infrared-safe observable that is non-zero at Born level is described at NLO accuracy.
Here, $\mathcal{F}^\mathrm{MEC}$ denotes the matrix-element corrected shower operator,
\begin{equation}
\begin{split}
    &\mathcal{F}^\mathrm{MEC}_0(O, \Phi_B) = \Delta_0(t_0,t_\mathrm{c})\, O(\Phi_B) \\
    &\qquad + \int\limits^{t_0}_{t_\mathrm{c}}\, \frac{R(\Phi_R)}{B(\Phi_B)}\, \Delta_0(t_0,t)\, \mathcal{F}_1(O, \Phi_{R})\, \D\Phi_{+1} \,,
\end{split}
\end{equation}
with all higher-multiplicity terms $\mathcal{F}_n$, $n>0$ identical to the uncorrected case. 
Any infrared-safe observables that are non-vanishing only starting from real-emission configurations are described at LO accuracy, while any observables related to higher-multiplicity configurations are only described by the parton shower.
However, the NLO accuracy of Born-level observables hinges on the preservation of tree-level kinematics in the shower evolution.
The matching strategy would fail to reproduce NLO-accurate predictions if additional radiation would change the underlying Born kinematics. 
In case of DIS, this would be the case if the parton-shower recoil scheme includes recoil against the scattering lepton. 

\section{Implementation in \pythia\ 8}
\label{sec:implementation}

We implement the NLO matching scheme in the \pythia\ event-generation framework, extending its internal
\begin{verbatim}
    WeakBosonExchange:ff2ff(t:gmZ)
\end{verbatim} 
process and relying on its internal parton-shower algorithms.
As NLO accuracy of our matching scheme rests upon the assumption that the parton shower leaves the Born-level variables $x$ and $Q^2$ unchanged, it can currently only be used with the ``dipole-recoil'' option of the simple shower or \vincia.
The method retains the NLO accuracy for other parton showers if and only if the shower preserves the Born variables and corrects the first branching to the real matrix element. 
Any parton shower with that changes the Born variables in the first branching or does not correct the first branching to the real matrix element breaks the NLO accuracy.

The implementation requires two separate steps, which intervene the event generation in \pythia\ at two distinct places:
\begin{enumerate}
  \item reweight Born-level events to the rate defined by $\bar{B}(\Phi_B)$, given in terms of Born kinematics;
  \item implement the matrix-element corrections to the parton shower to achieve the correct branching rate of the real matrix element $R(\Phi_{R})$.
\end{enumerate}
Although the $\bar{B}$-factor is process-dependent, the same matching is extensible to any parton shower which leaves the Born kinematics unaltered and generates the correct rate of the first branching.
The NLO event reweighting is done at an early level in the event-generation chain, so as to include it in the unweighting.
This way, events are generated with unit weight and there is no need for an additional unweighting procedure.
Nevertheless, users can access the NLO weight by calling the function
\begin{verbatim}
    double Info::weightNLO()
\end{verbatim}
to fetch the multiplicative NLO correction factor.
This functionality can be applied to study the effect from the NLO correction for given kinematics.

The NLO weight can be calculated for each incoming flavor separately, or by considering all flavor channels and summing over them. We adopt the latter method and perform flavor selection as in the LO case, based on the relative couplings and PDFs. In this way, the reweighting can take advantage of the existing infrastructure. 
The $Z$ boson contribution induces an opposite-sign term for quarks and anti-quarks, which would currently not be captured by this algorithm. These effects are, however, only resolved by flavor-tagged observables, for which our algorithm may be extended in the future. See, e.g. \cite{vanBeekveld:2025lpz} for an account of their importance. Our implementation provides NLO accuracy for flavor-summed observables, and is compatible with the existing switch \texttt{WeakZ0:gmZmode} which can be used to study different contributions one-by-one.
We note that also the exact form of matrix-element corrections, discussed below, may impact the accuracy of the matching scheme for flavor-tagged observables, especially in the case of flavor-changing initial-state branchings \cite{vanBeekveld:2025lpz}.

While all LO and NLO expressions consider massless partons and leptons only, phase-space points are shifted on-shell afterwards, following standard \pythia\ event-generation procedures.
In case of DIS, construction of the kinematics has been updated in version 8.316 such that $Q^2$ remains intact when finite masses are assigned to particles. 

\paragraph{Matrix-element corrections} for the DIS process can be obtained as ratios of fixed-order matrix elements and parton-shower rates
\begin{equation}
    R_{\gamma^*q\to qg}(z, z_1) = \frac{\mathrm{d}\sigma^\mathrm{ME}_{\gamma^*q\to qg} / \mathrm{d}z \mathrm{d}z_1}{\mathrm{d}\sigma^\mathrm{PS}_{\gamma^*q\to qg} / \mathrm{d}z \mathrm{d}z_1} 
\end{equation}
\begin{equation}
    R_{\gamma^*g\to q\bar{q}}(z, z_1) = \frac{\mathrm{d}\sigma^\mathrm{ME}_{\gamma^*g\to q\bar{q}} / \mathrm{d}z \mathrm{d}z_1}{\mathrm{d}\sigma^\mathrm{PS}_{\gamma^*g\to q\bar{q}} / \mathrm{d}z \mathrm{d}z_1},
\end{equation}
where $z_1 = p \cdot p_1 / p\cdot q $. Contributions from MECs to $Z^0 q \to qg$ and $Z^0 g \to q \bar{q}$ as well as interferences between photon- and $Z$-induced amplitudes are expected to be negligible, and are not considered here. These effects may be considered in the future, either through explicit $eq \to eqg$ and $eg \to eq\bar{q}$ MECs or via multi-jet merging techniques \cite{Helenius:2024wjg}. This would give a consistent NLO description of NC DIS over the full $Q^2$ range. 

The real-emission matrix-element rates are obtained from the amplitudes in Eq.~\eqref{eq:ME_Rq} and \eqref{eq:ME_Rg} in $D=4$ dimensions by taking the limit $\varepsilon \to 0$ and using the relations 
\begin{equation}
    \quad \hat{s} = \frac{Q^2}{z}(1-z)\,, \quad \hat{t} = -\frac{Q^2}{z}(1-z_1) \,, \quad \hat{u} = -\frac{Q^2}{z}z_1 \,,
\end{equation}
which gives the rates 
\begin{equation}
    \label{eq:MErate_Rq}
    \mathrm{d}\sigma^\mathrm{ME}_{\gamma^*q\to qg} = \sigma_0 \frac{\alphaS}{2\pi} \CF \bigg[\frac{z^2+z_1^2}{(1-z)(1-z_1)} + 2(1+3z z_1) \bigg] \mathrm{d}z \mathrm{d}z_1
\end{equation}
and 
\begin{equation}
\begin{split}
    \mathrm{d}\sigma^\mathrm{ME}_{\gamma^*g\to q\bar{q}} = \sigma_0 \frac{\alphaS}{2\pi} \TR \bigg[ &\frac{(z_1^2 + (1-z_1)^2)(z^2 + (1-z)^2)}{z_1(1-z_1)} \\
    + &8z(1-z) \bigg] \mathrm{d}z \mathrm{d}z_1 \,,
\end{split}
\label{eq:MErate_Rg}
\end{equation}
where $\sigma_0$ is the common LO factor across all matrix elements. The rates match the matrix elements presented in the appendix of \cite{Catani:1996vz}. 
For the gluon-initiated process, the full matrix-element contains both cases, where either the quark or the antiquark evolves backwards into the gluon, corresponding to two distinct parton-shower histories.
In order to yield a finite correction for the branching rate, we rewrite this matrix element by partial fractioning the first term in Eq.~\eqref{eq:MErate_Rg} and dividing the finite term in the second line equally between both histories,
\begin{equation}
\begin{split}
    \mathrm{d}\sigma^\mathrm{ME}_{\gamma^*g\to q\bar{q}} = \sigma_0 \frac{\alphaS}{2\pi} \TR \Bigg[ \frac{z_1(z^2+(1-z)^2)}{1-z_1} + &4z(1-z) \\
    + \frac{(1-z_1)(z^2+(1-z)^2)}{z_1} + &4z(1-z) \Bigg] \mathrlap{{} \mathrm{d}z \mathrm{d}z_1 \,,}
\end{split}
\label{eq:MErate_Rg_div}
\end{equation}
where the first line corresponds to a quark and the latter to an antiquark coupling to the photon and the sum over both shower histories yields the same expression as in Eq.~\eqref{eq:MErate_Rg}.
The parton-shower emission rates are defined as
\begin{equation}
    \label{eq:PSrate}
    \mathrm{d}\sigma^\mathrm{PS}_{\gamma^*a\to bc} = \sigma_0\, K_{a\to bc}(t,z)\, \D t \D z \,.
\end{equation}

\begin{figure*}
    \centering
    \begin{subfigure}{0.45\textwidth}
        \centering
        \includegraphics[width=\linewidth]{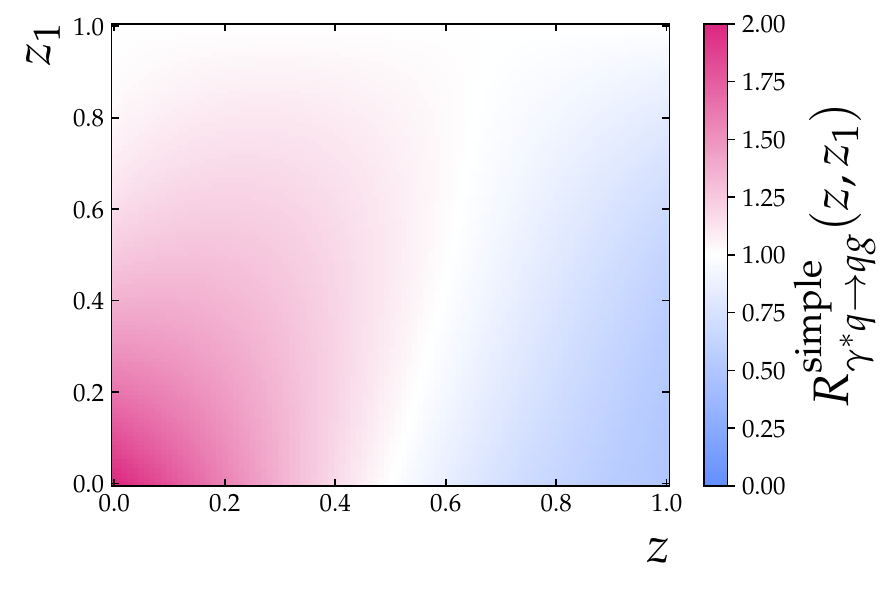}
        \caption{}
    \end{subfigure}
    \begin{subfigure}{0.45\textwidth}
        \centering
        \includegraphics[width=\linewidth]{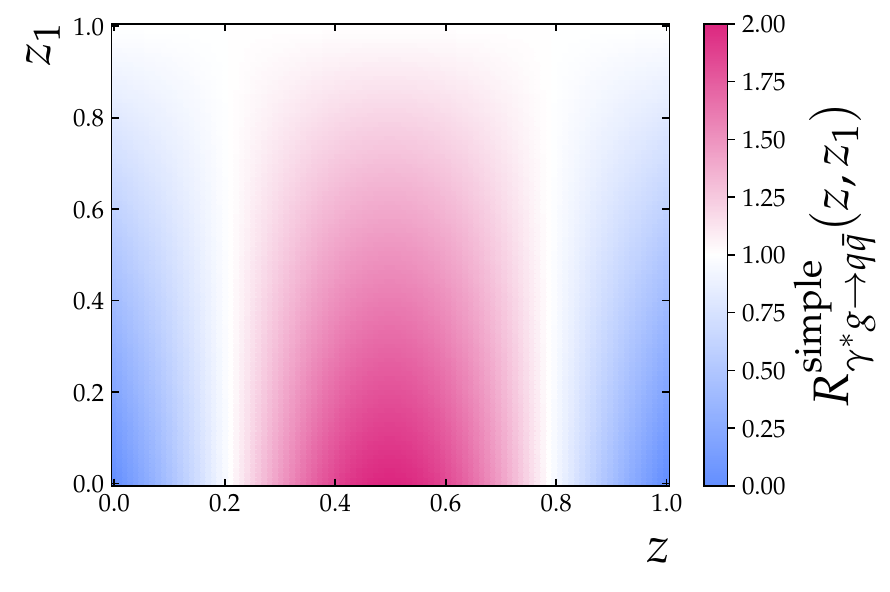}
        \caption{}
    \end{subfigure}
    \begin{subfigure}{0.45\textwidth}
        \centering
        \includegraphics[width=\linewidth]{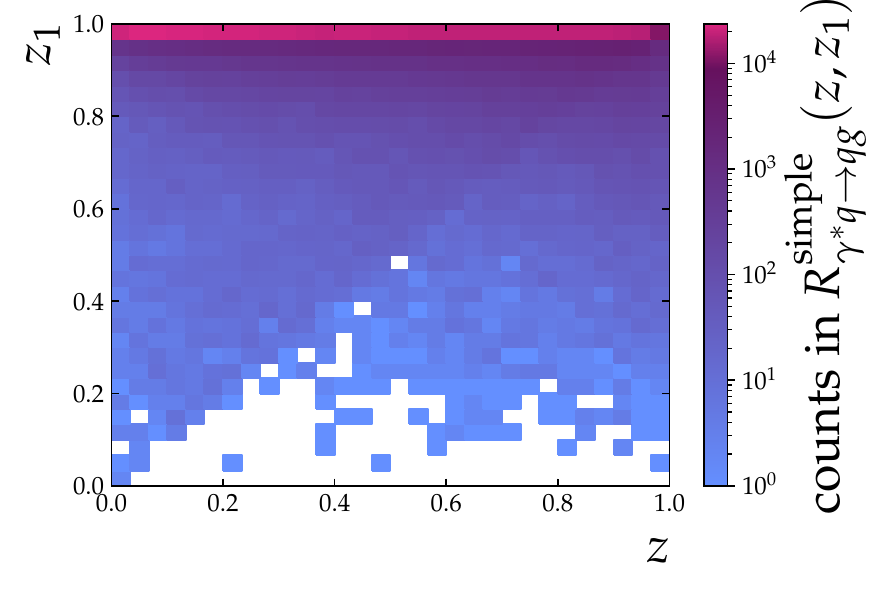}
        \caption{}
    \end{subfigure}
    \begin{subfigure}{0.45\textwidth}
        \centering
        \includegraphics[width=\linewidth]{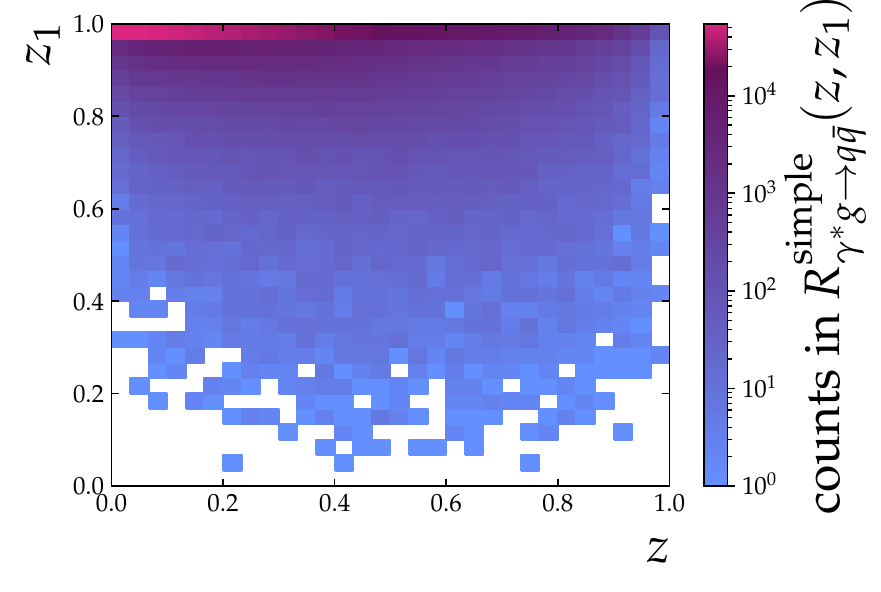}
        \caption{}
    \end{subfigure}
    \caption{Matrix-element corrections of DIS $\gamma^* a \to bc$ processes for the default shower in \pythia. Top figures (a) and (b) show magnitudes of corrections, and bottom figures (c) and (d) show event counts from 2 million generated DIS events. }
    \label{fig:MEC_phasespace}
\end{figure*}

\begin{figure}
    \centering
    \includegraphics[width=\linewidth]{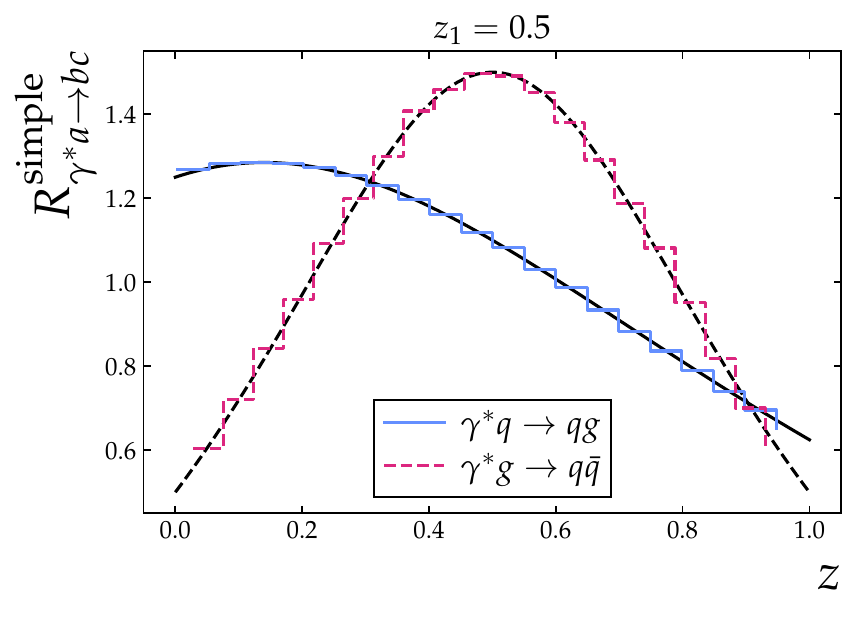}
    \caption{A sample of simple-shower matrix-element corrections at $z_1 = 0.5$. Solid lines are the analytic expressions of Eqs. \eqref{eq:MEC_q} and \eqref{eq:MEC_g}. Dashed histograms are corrections in the range $z_1 \in [0.49, 0.51]$, sampled from hardest emissions in $2$ million \pythia\ DIS events. }
    \label{fig:MEC_phasespace_slice}
\end{figure}

\paragraph{For the simple shower} the branching rates can be expressed in terms of $z$ and $z_1$ as in \cite{Cabouat:2017rzi} with
\begin{align}
         &t = (1-z)(-m_{b}^2) = \frac{(1-z)(1-z_1)}{z}m_\mathrm{dip}^2, \\
         &\frac{1}{t} \, \mathrm{d}t \mathrm{d}z = \frac{\mathrm{d}z \mathrm{dz_1}}{1-z_1} \,,
\end{align}
where $m_\mathrm{dip}$ is the invariant mass of the color dipole, giving
\begin{equation}
    \label{eq:PSrate_Rq}
    \mathrm{d}\sigma^\mathrm{simple}_{\gamma^*q\to qg} = \sigma_0 \frac{\alphaS}{2\pi} \CF \frac{1+z^2}{(1-z)(1-z_1)} \mathrm{d}z \mathrm{d}z_1\,,
\end{equation}
and 
\begin{equation}
    \label{eq:PSrate_Rg}
    \mathrm{d}\sigma^\mathrm{simple}_{\gamma^*g\to q\bar{q}} = \sigma_0 \frac{\alphaS}{2\pi} \TR \frac{z^2 + (1-z)^2}{1-z_1} \mathrm{d}z \mathrm{d}z_1\,. 
\end{equation} 
Using these rates, the matrix-element corrections are written as
\begin{equation}
    \label{eq:MEC_q}
    R_{\gamma^*q\to qg}^\mathrm{simple}(z, z_1) = \frac{z^2 + z_1^2 + 2(1+3z_1z)(1-z)(1-z_1)}{1+z^2}\,,
\end{equation}
and 
\begin{equation}
    \label{eq:MEC_g}
    R_{\gamma^*g\to q\bar{q}}^\mathrm{simple}(z, z_1) = z_1 + \frac{4z(1-z)(1-z_1)}{z^2 + (1-z)^2}\,.
\end{equation}
Magnitudes of these corrections are plotted in terms of $z$ and $z_1$ for simple shower in Figs. \ref{fig:MEC_phasespace} (a) and (b). 
In the soft $(z=1, z_1=1)$ and collinear $(z_1=1, m_b^2=0)$ limits, both are equal to $1$. The correction of Eq.~\eqref{eq:MEC_q} is the same for quarks and anti-quarks. 
These corrections are essential for the matching strategy, as stated in section \ref{sec:matching}. However, they have little effect for the space-like shower used in DIS processes, since the shower rates of Eq.~\eqref{eq:PSrate_Rq} and \eqref{eq:PSrate_Rg} closely resemble the structure of real-emission matrix elements of Eq.~\eqref{eq:ME_Rq} and \eqref{eq:ME_Rg}. Moreover, Figs. \ref{fig:MEC_phasespace} (c) and (d) show that the majority of first emissions occur at $z_1\approx 1$, where the corrections are close to unity. In Fig. \ref{fig:MEC_phasespace_slice}, fixed-order matrix-element corrections of Eqs. \eqref{eq:MEC_q} and \eqref{eq:MEC_g} are compared with MECs from first emissions near $z_1=0.5$ in DIS events. We find agreement between the hardest emissions and the real-emission matrix elements, which holds true across the whole $z_1$ range.

\paragraph{For \textsc{Vincia}} the parton-shower rates can be written in terms of $z$ and $z_1$ by substituting
\begin{equation}
    z = 1-y_{jk}\,, \quad z_1 = 1-y_{ij}\,,
\end{equation}
so that the shower rates are given by
\begin{align}
    \mathrm{d}\sigma^\mathrm{\textsc{Vincia}}_{\gamma^*q\to qg} &= \sigma_0 \frac{\alphaS}{2\pi} \CF \frac{(1+z^2)(1+z_1^2)}{2(1-z)(1-z_1)} \mathrm{d}z \mathrm{d}z_1\,,     \label{eq:VinciaRate_Rq} \\
    \mathrm{d}\sigma^\mathrm{\textsc{Vincia}}_{\gamma^*g\to q\bar{q}} &= \sigma_0 \frac{\alphaS}{2\pi} \TR \frac{z^2+(1-z)^2}{1-z_1} \mathrm{d}z \mathrm{d}z_1\,. \label{eq:VinciaRate_Rg}
\end{align}
This yields the matrix-element corrections 
\begin{equation}
\begin{split}
    R_{\gamma^*q\to qg}^\mathrm{\textsc{Vincia}}(z, z_1) &= \frac{2(z_1^2+z^2)}{(1+z_1^2)(1+z^2)} \\
    &+ (1-z_1)\frac{4(1-z)(1+3z z_1)}{(1+z^2)(1+z_1^2)} \,,
\end{split}
\label{eq:MEC_q_vincia}
\end{equation}
and 
\begin{equation}
    R_{\gamma^*g\to q\bar{q}}^\mathrm{\textsc{Vincia}}(z, z_1) = z_1 + \frac{4z(1-z)(1-z_1)}{z^2 + (1-z)^2}\,.
\label{eq:MEC_g_vincia}
\end{equation}
As for the simple shower, these are finite everywhere and approach unity in the soft and collinear limits. Notice that the shower rate for the gluon-initiated branching, Eq.~\eqref{eq:VinciaRate_Rg}, is equal to the one in simple shower, Eq.~\eqref{eq:PSrate_Rg}, and thus the corrections in Eqs.~\eqref{eq:MEC_g} and \eqref{eq:MEC_g_vincia} have the same expressions. The MECs are applied for DIS by default since \pythia\ version 8.317 independently from NLO matching.

\section{Validation}
\label{sec:validation}

For the validation of the matching strategy, we compare it with predictions from \sherpa\ fixed-order DIS at NLO \cite{Schumann:2007mg,Krauss:2001iv,Gleisberg:2008fv,Gleisberg:2007md,Chahal:2022rid,Bothmann:2016nao} and the DIS module of \powhegbox\ \cite{Banfi:2023mhz} at the level of the hard process. These methods also produce NLO-accurate predictions of inclusive quantities, so a natural cross-check is to compare the distributions of the DIS quantities $x$, $y$, $Q^2$ and $W^2$. We use a fixed value for the electromagnetic coupling with $1 / \alpha = 137.036$. We derive the electroweak parameters from the input of $\alpha(M_Z)$, electroweak mixing angle $\sin^2 \theta_W = 0.2312$, and the $Z$-boson mass $M_Z = 91.1876$ GeV. We use the PDF set \texttt{NNPDF31\_nlo\_as\_0118\_luxqed} \cite{NNPDF:2017mvq} provided by the LHAPDF library \cite{Buckley:2014ana}. The running of the strong coupling is handled by the PDF set. The recommended \pythia\ setup and parameters for NLO DIS event generation are presented in Table \ref{tab:settings}. The recommended settings for simple-shower option are also listed in the example main program \texttt{main341.cc}, supplied in the examples directory in \pythia\ releases.

\begin{figure*}[h]
     \centering
     \begin{subfigure}[b]{0.45\textwidth}
         \centering
         \includegraphics[width=\textwidth]{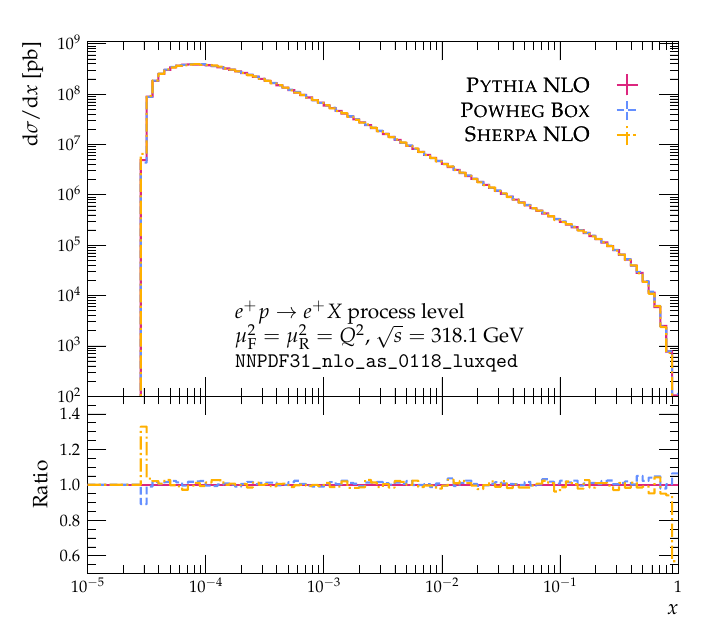}
         \caption{}
     \end{subfigure}
     \begin{subfigure}[b]{0.45\textwidth}
         \centering
         \includegraphics[width=\textwidth]{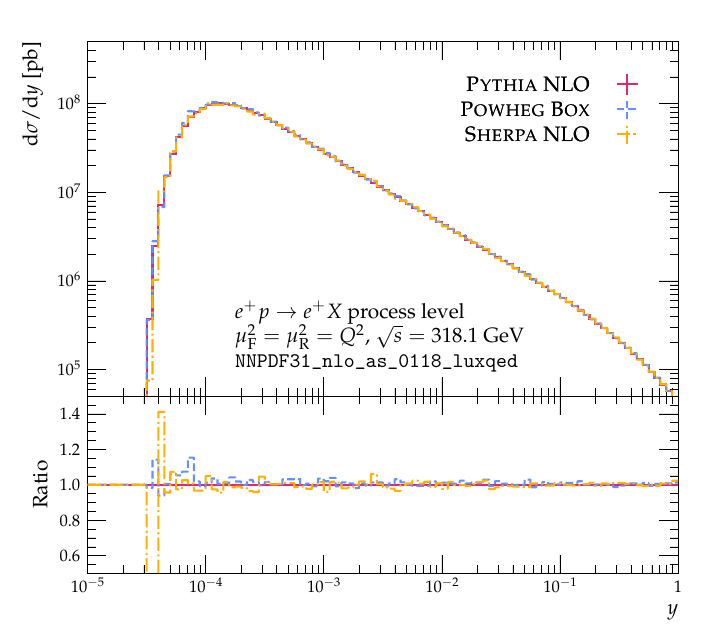}
         \caption{}
     \end{subfigure}
     \hfill
     \begin{subfigure}[b]{0.45\textwidth}
         \centering
         \includegraphics[width=\textwidth]{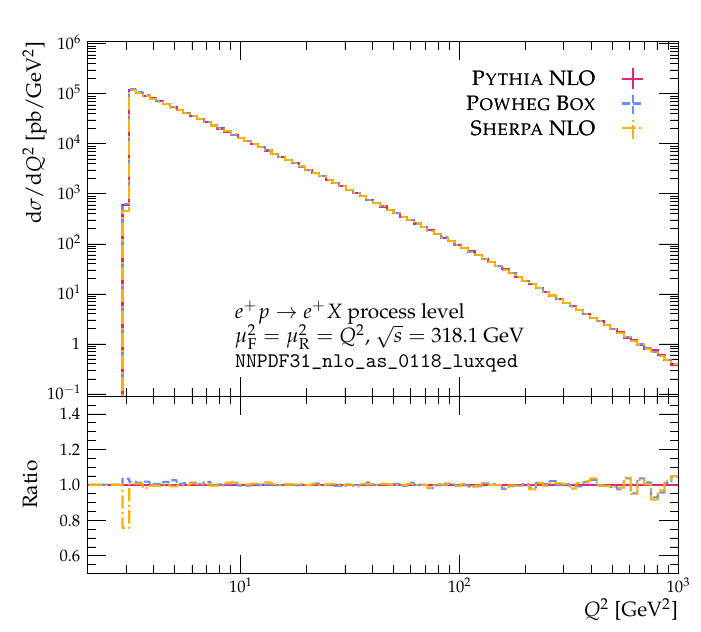}
         \caption{}
     \end{subfigure}
     \begin{subfigure}[b]{0.45\textwidth}
         \centering
         \includegraphics[width=\textwidth]{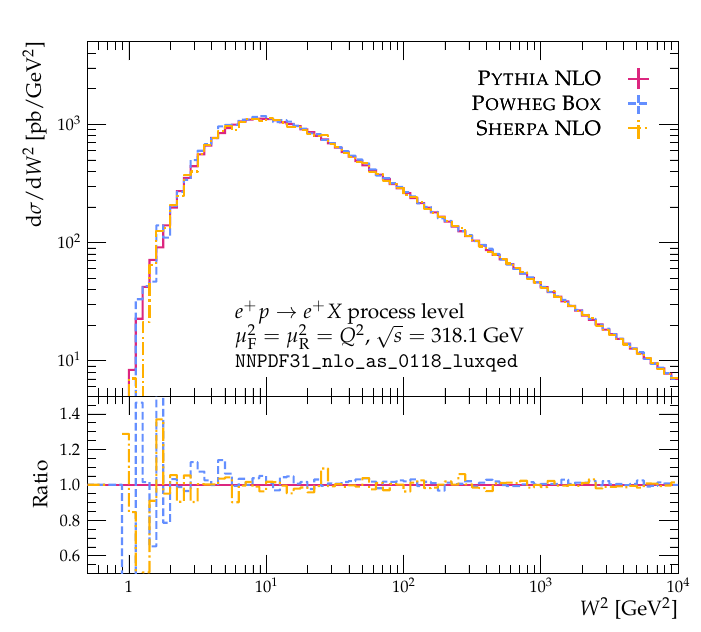}
         \caption{}
     \end{subfigure}
        \caption{Differential cross sections in terms of a) momentum fraction $x$, b) inelasticity $y$, c) virtuality $Q^2$, and d) invariant hadronic mass $W^2$ at NLO as predicted by \pythia, \powhegbox\ and \sherpa. Only the hard process is generated, with no additional effects from beam remnants, parton showers or hadronization. }
        \label{fig:DISinvariants}
\end{figure*}

Setting up event generation with beam energies corresponding to the HERA experiments, $27.5$ GeV for the electron and $920$ GeV for the proton, we generate events with the three generators at process level; without parton showers, beam remnants, or hadronization. 

To focus solely on the higher-order effects, all partons are treated as massless, and the electron is treated as point-like, with no electron PDF, no QED emissions present, and a fixed electromagnetic coupling. We impose a lower cut in propagator virtuality at $Q^2 = 3.1$ GeV$^2$ to regulate the cross section and stay within the region of DIS. The cross sections differential in DIS quantities $x$, $y$, $Q^2$ and $W^2$ are shown in Fig.~\ref{fig:DISinvariants}. Although the implementation of NLO calculations differ among the three approaches, the agreement between the predictions is valid across a large kinematical domain, spanning multiple orders of magnitude in cross section. 

\begin{table*}
\centering
\begin{tabular}{||l| c l c c||} 
 \hline
 Setting & Recommended for NLO DIS & Parameter & Value & Default \\ [0.5ex] 
 \hline\hline
 $\alphaS$ value & 0.118 & \texttt{SigmaProcess:alphaSvalue} & \texttt{0.118} & \texttt{0.130} \\
 \hline 
 $\alphaS$ running & 2nd order & \texttt{SigmaProcess:alphaSorder} & \texttt{2} & \texttt{1} \\
 \hline
 Factorization scale & $Q^2$ & \texttt{SigmaProcess:factorScale2} & \texttt{6} & \texttt{1} \\
 \hline
 Renormalization scale & $Q^2$ & \texttt{SigmaProcess:renormScale2} & \texttt{6} & \texttt{1} \\
 \hline 
 PDF set & \texttt{NNPDF31\_nlo\_as\_0118\_luxqed} & \texttt{PDF:pSet} & \texttt{19} & \texttt{13} \\
 \hline
 Electron PDF & off & \texttt{PDF:lepton} & \texttt{off} & \texttt{on} \\ 
 \hline\hline
 \multicolumn{5}{||l||}{Simple shower} \\ 
 \hline
 Shower starting scale & $s$ & \texttt{SpaceShower:pTmaxMatch} & \texttt{2} & \texttt{0} \\
 \hline
 QED emissions off electron & off & \texttt{TimeShower:QEDshowerByL} & \texttt{off} & \texttt{on} \\
 \hline
 Recoil scheme & Dipole recoil & \texttt{SpaceShower:dipoleRecoil} & \texttt{on} & \texttt{off} \\ 
 \hline\hline
 \multicolumn{5}{||l||}{\vincia} \\
 \hline
 Shower starting scale & $s$ & \texttt{Vincia:pTmaxMatch} & \texttt{2} & \texttt{0} \\
 \hline
 QED emissions & off & \texttt{Vincia:ewMode} & \texttt{0} & \texttt{2} \\
 \hline
 Matrix-element corrections & on & \texttt{Vincia:modeMECs} & \texttt{1} & \texttt{0} \\
 ~ & ~ &  \texttt{Vincia:useInternalMEs} & \texttt{on} & \texttt{off}\\
 ~ & ~ & \texttt{Vincia:maxMECs2to2} & \texttt{1} & \texttt{0} \\
 [1ex] 
 \hline
\end{tabular}
\caption{Recommended settings for DIS event generation with NLO+PS matching in \pythia\ 8. }
\label{tab:settings}
\end{table*}

\begin{figure}
    \centering
    \includegraphics[width=0.5\textwidth]{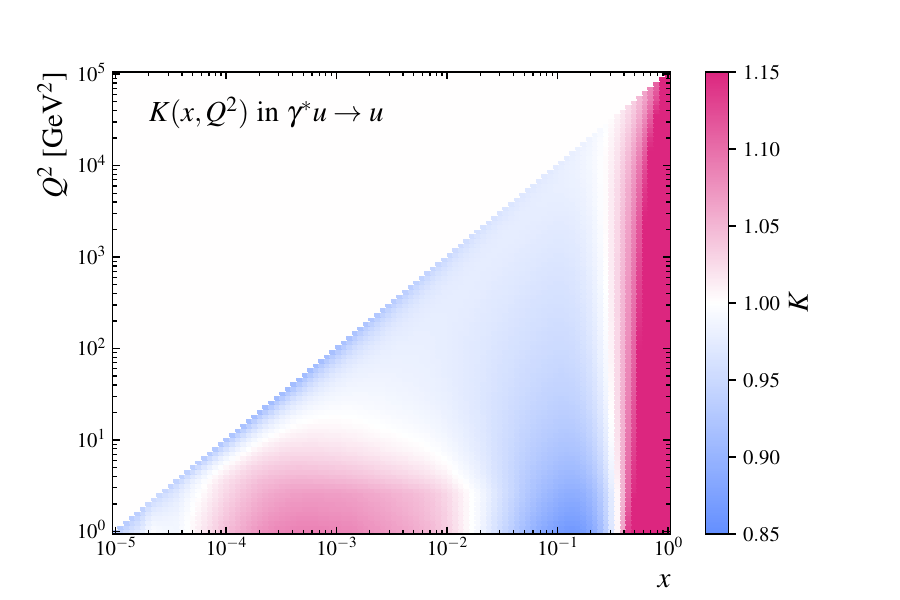}
    \caption{The NLO correction to the electron-proton cross section for an incoming $u$-quark. }
    \label{fig:Heatmap_K}
\end{figure}

The MC@NLO matching strategy is subtractive by nature, containing subtraction terms similar to NLO-subtraction schemes, and generates both Born-like S-events and real-emission H-events. These events may come with negative weights. The multiplicative matching presented in this work, similar to POWHEG-style matching, comes with a Born-local $K$-factor that reweights the Born-level events. The value of this $K$-factor is highlighted in a grid of $x$ and $Q^2$ for the subprocess $\gamma^* u \to u$ in Fig.~\ref{fig:Heatmap_K}. Notably, the $K$-factor is larger than $1$ at small values of $Q^2$ when $10^{-4}<x<10^{-2}$ and at $x>0.4$ but below $1$ in other regions.

\begin{figure*}
     \centering
     \begin{subfigure}[b]{0.45\textwidth}
         \centering
         \includegraphics[width=\textwidth]{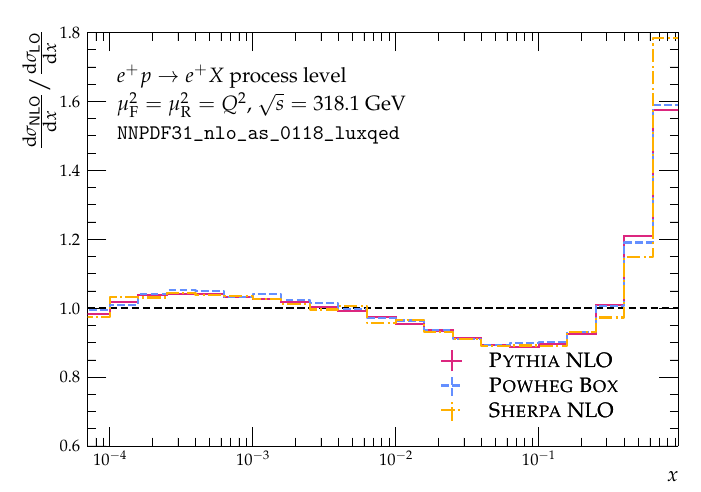}
         \caption{}
     \end{subfigure}
     \begin{subfigure}[b]{0.45\textwidth}
         \centering
         \includegraphics[width=\textwidth]{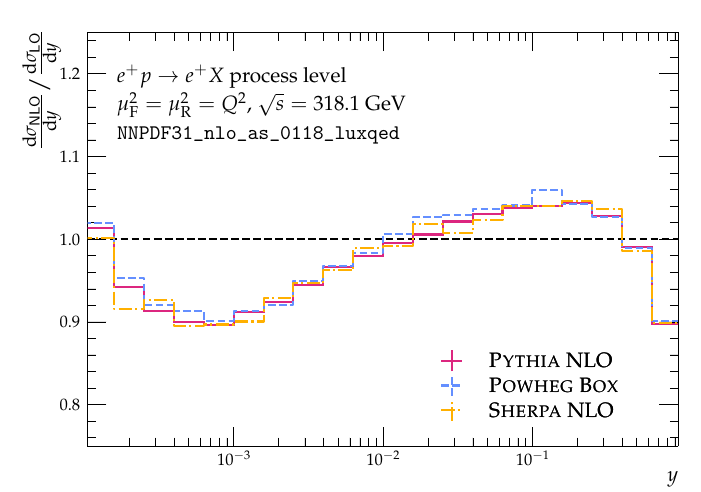}
         \caption{}
     \end{subfigure}
     \hfill
     \begin{subfigure}[b]{0.45\textwidth}
         \centering
         \includegraphics[width=\textwidth]{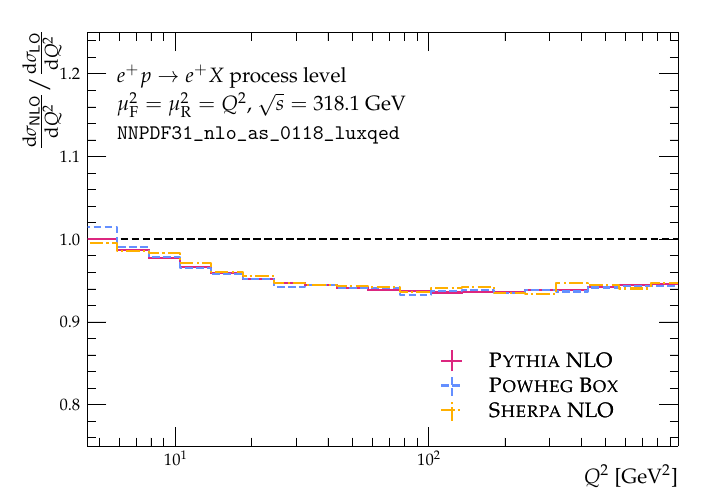}
         \caption{}
     \end{subfigure}
     \begin{subfigure}[b]{0.45\textwidth}
         \centering
         \includegraphics[width=\textwidth]{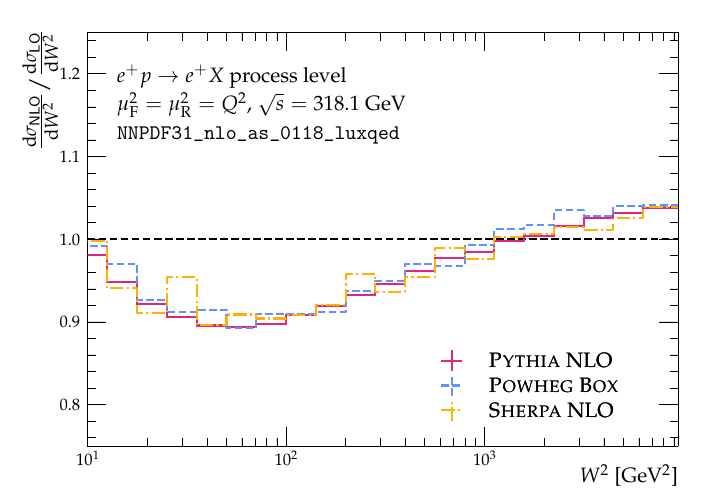}
         \caption{}
     \end{subfigure}
        \caption{Comparisons of the ratios of NLO to LO hard-process cross-sections, differential in a) momentum fraction $x$, b) inelasticity $y$, c) virtuality $Q^2$, and d) invariant hadronic mass $W^2$ from \pythia, \powhegbox\ and \sherpa.}
        \label{fig:Kfactors}
\end{figure*}

We compare the NLO $K$-factors of the three generators, again differential in the DIS variables, in Fig.~\ref{fig:Kfactors}. The $K$-factor here is defined as the ratio of the differential NLO and LO cross sections, displaying the impact of NLO corrections. The total integrated DIS cross sections are only slightly decreased at NLO, with a difference of less than $2\%$, while the influence on differential cross sections is around $10\%$ in certain regions of the phase space. We again find a very good agreement between the three generators.

\section{Comparisons to HERA data}
\label{sec:results}

Measurements of reduced cross-sections at the HERA collider \cite{H1:2015ubc}, performed by the H1 and ZEUS collaborations, offer high-precision data over a wide kinematic range in the $x$ variable and virtuality $Q^2$ in numerous center-of-mass energies. These measurements are directly sensitive to the structure functions, and provide a suitable setting to test the effects of higher-order corrections with the presented strategy. 

\begin{figure*}
     \centering
     \begin{subfigure}[b]{0.45\textwidth}
         \centering
         \includegraphics[width=\textwidth]{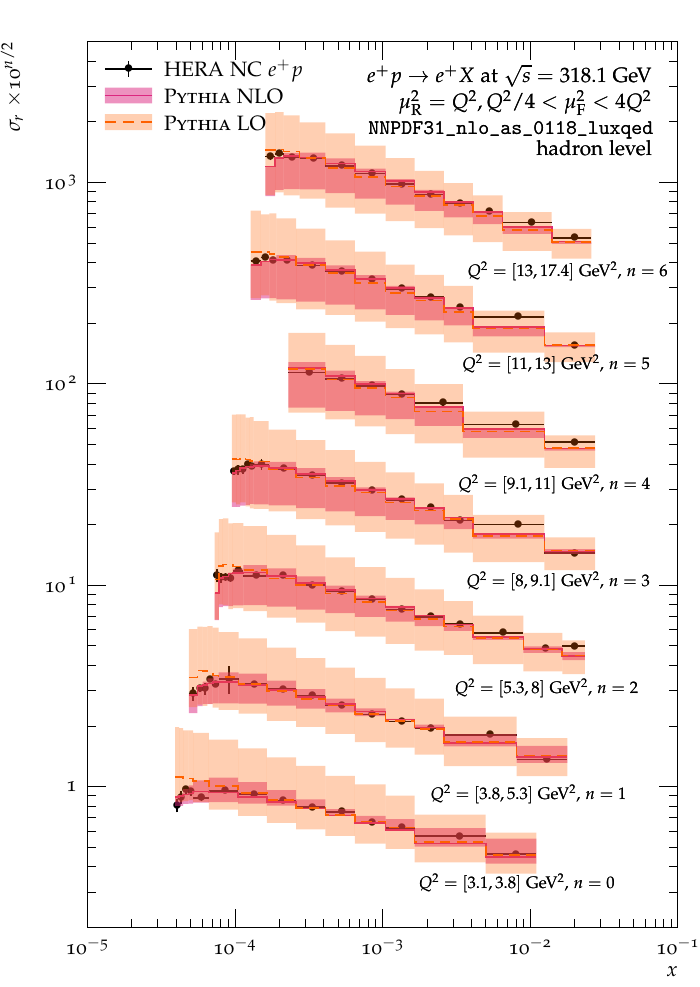}
     \end{subfigure}
     \hfill
     \begin{subfigure}[b]{0.45\textwidth}
         \centering
         \includegraphics[width=\textwidth]{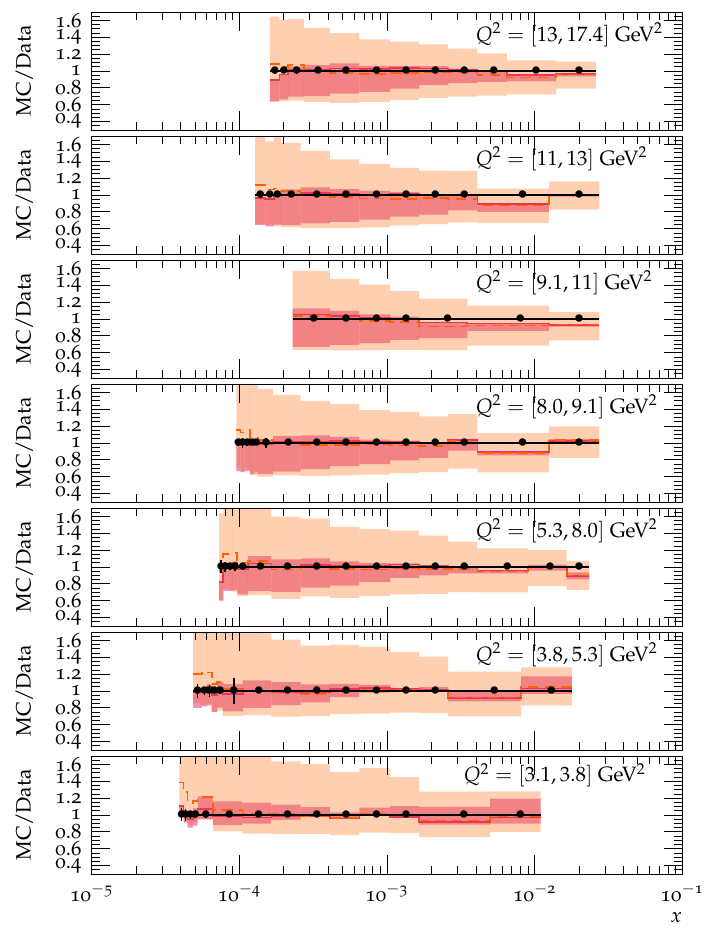}
     \end{subfigure}
        \caption{The reduced positron-proton cross-sections in unpolarized neutral-current deep inelastic scattering, as a function of $x$. The hadron-level predictions from \pythia\ are shown at leading order (dashed line) and NLO+PS matched next-to-leading order (solid line), with the $3$-point factorization scale variations shown in the envelopes. The data points are from the combined H1 and ZEUS measurements \cite{H1:2015ubc}. }
        \label{fig:HERA1}
\end{figure*}

The data set for unpolarized neutral and charged-current cross sections is available in the routine HERA\_2015\_I1377206 of Rivet analysis toolkit \cite{Bierlich:2019rhm}. The reduced cross section is given by 
\begin{equation}
    \sigma_r = \frac{\D^2 \sigma}{\D x\D Q^2} \frac{xQ^4}{2\pi\alpha^2 Y_+} = F_2 - \frac{y^2}{Y_+}F_L \mp \frac{Y_-}{Y_+}xF_3,
\end{equation}
where $Y_\pm = 1 \pm (1-y)^2$. For these data comparisons, we employ the full event simulation at leading-order and matched next-to-leading order, with beam remnants, parton showers and hadronization effects, as opposed to the process-level comparisons of section \ref{sec:validation}. We use the same electroweak settings, scale choices and PDF sets. Here, we only show results with the dipole-recoil scheme of the simple shower as the results with \vincia\ were practically identical. This is, of course, expected since the parton shower is not supposed to modify scattered lepton kinematics from which $x$ and $Q^2$ are derived. The reduced cross-sections for positron-proton scattering at $s = 318$ GeV and $Q^2$ values between $3.1$ GeV$^2$ and $17.4$ GeV$^2$ are presented in Fig.~\ref{fig:HERA1}. The missing effects from higher-order perturbation theory are estimated by varying the factorization scale with factors of $0.5$ and $2$. These variations are shown as envelopes. We vary the factorization scale only, since at leading order, the hard process is not sensitive to variations of the renormalization scale; these variations have an effect through the running of the strong coupling, which becomes relevant only at next-to-leading order. 

The leading-order description agrees well with the data, but suffers from large uncertainties through the scale variations. These variations are significantly reduced with the NLO matching strategy. The reduced cross-sections at small values of $x$ have considerable improvements compared to the LO description. 

The comparisons at higher virtualities are included in the Appendix, in figures \ref{fig:HERA2}-\ref{fig:HERA5}. It is evident that the effects of next-to-leading order corrections are less relevant on higher virtualities, and we see good agreement also with a leading-order prediction. 
Overall, we find good agreement between the data and the NLO matched prediction throughout the whole kinematic range which confirms that the implementation retains the expected fixed-order accuracy throughout the whole event-generation chain.

\section{Applications to the EIC}
Finally, we apply our implementation to produce preliminary predictions for the EIC \cite{AbdulKhalek:2021gbh}.
To this end, we consider electron-proton collisions with an electron beam energy of $E_e = 18~\text{GeV}$ and a proton beam energy of $E_p = 275~\text{GeV}$.
We use the \texttt{NNPDF31\_nlo\_as\_0118\_luxqed} PDF set \cite{NNPDF:2017mvq} with second-order running strong coupling.
Electroweak parameters are set according to the default values in \pythia,
\begin{equation}
\begin{split}
    &\alpha(M_Z) = \frac{1}{127.92}\,,\quad M_Z = 91.1876~\text{GeV}\,,\\
    &\sin^2\theta_\mathrm{W} = 0.2312\,,
\end{split}
\end{equation}
with first-order running of the electromagnetic coupling.
Jets are defined in the anti-$k_\mathrm{T}$ algorithm with $R = 0.4$ using FastJet \cite{Cacciari:2011ma}.
We focus on the region
\begin{equation}
    Q^2 > 100~\text{GeV}^2\,,
\end{equation}
so as to focus on the part of phase space which is less susceptible to higher-multiplicity final states \cite{Carli:2010cg,Helenius:2024wjg}.
We place the following cuts on reconstructed jets in the lab frame
\begin{equation}
    p_\mathrm{T} > 5~\text{GeV}\,, \quad \vert\eta\vert < 4\,.
\end{equation}

\begin{figure*}
    \centering
    \begin{subfigure}[b]{0.45\textwidth}
    \centering
    \includegraphics[width=\textwidth]{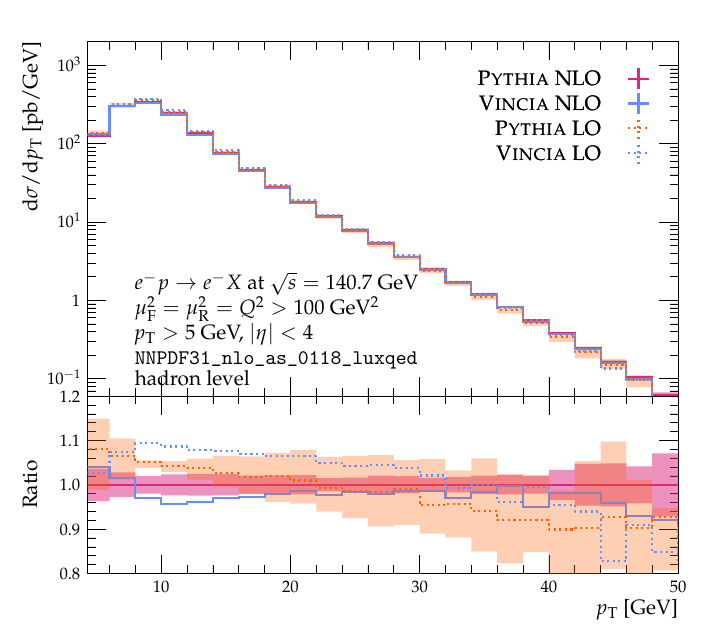}
    \end{subfigure}
    \hfill
    \begin{subfigure}[b]{0.45\textwidth}
    \centering
    \includegraphics[width=\textwidth]{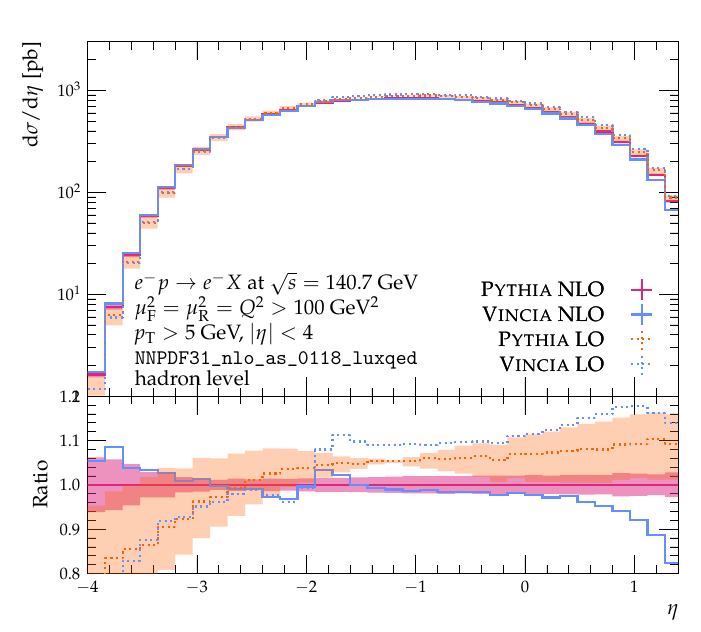}
    \end{subfigure}
    \caption{\pythia\ and \vincia\ predictions at LO+PS and NLO+PS for the exclusive transverse momentum (left) and rapidity (right) distributions of the first jet in the lab frame for EIC kinematics. Scale variations are shown as colored bands, considered only with the \pythia\ results. For the LO result, the factorization scale is varied from $\frac{1}{2}\MUF$ to $2\MUF$, and for the NLO result the renormalization scale is simultaneously varied. }
    \label{fig:EIC}
\end{figure*}

Fig.~\ref{fig:EIC} shows the exclusive transverse momentum and rapidity distributions of the first jet in the lab frame.
For the \pythia\ default shower we vary renormalization and factorization scales in the NLO calculation (7-point variation) but only the latter in LO case (3-point variation).
The scale variations are significantly reduced with the NLO matched calculation and the resulting uncertainty bands mostly coincide with the LO variations. We focus here solely on scale variations to which the implemented NLO matching has a direct impact but acknowledge that further modeling uncertainties related shower-scale variations and hadronization exists. 
While we observe differences in the \pythia\ and \vincia\ predictions at LO (dotted lines) over the entire $p_\mathrm{T}$ spectrum and for $\eta > -2$, we find very good agreement between the two shower algorithms after the matching to NLO (solid lines).
Except for the very forward region $\eta > 0.5$, the \pythia\ and \vincia\ predictions agree with each other within 5-10\%.
The larger effect from the NLO matching with \vincia\ is due to larger MECs as the applied splitting kernels are not as close to the matrix elements as in case of the default \pythia\ shower as can be seen from the derived corrections in Eqs.~(\ref{eq:MEC_q}) and (\ref{eq:MEC_q_vincia}) for quark-initiated contribution and Eqs.~(\ref{eq:MEC_g}) and (\ref{eq:MEC_g_vincia}) for gluon-initiated one.
Over the bulk of the spectrum, the \vincia\ prediction further lies near the border of the seven-point scale variation band obtained at fixed order for the \pythia\ prediction, i.e., without taking into account scale variations in the shower itself.
We note that we find slightly larger disagreement between the two showers here compared to \cite{Banfi:2023mhz}.
We attribute this to the fact that in their case, the first branching is handled by \powhegbox\ and only subsequent branchings are handled by either \pythia\ or \vincia, while in our case, even the first branching is handled by different showers, using different evolution schemes.

\section{Conclusion}
We have presented next-to-leading order accurate event generation with a multiplicative parton-shower matching strategy for deep-inelastic scattering in the \pythia\ event generator. The setup has been implemented for neutral-current charged-lepton DIS processes including matrix-element corrections for the first parton-shower branching. 
We implement the matching strategy by calculating an NLO/LO weight using the inclusive NC DIS cross section and apply this weight to each event sampled according to the Born-level kinematics. In the weight calculation we utilize a numerical integration to obtain the necessary convolutions between PDFs and NLO coefficient functions in the adopted structure-function approach.
The real-emission correction factors in the NLO calculation now accounted in the PS kernels ensure that the first emission is generated according to the matrix elements. While such corrections have been in place for many PS-splittings in \pythia, this has not been the case for DIS. The algorithm is generic and can be applied to any parton-shower implementation which keeps the Born phase-space variables, i.e. the scattered lepton momentum, unchanged. We have made this available in \pythia\ version 8.317 for the default parton shower, in which the dipole-recoil variant is applicable to DIS, and for \vincia\, with further extension to e.g. \textsc{Apollo} parton shower \cite{Preuss:2024vyu} foreseen in the future. The use of this matching strategy is demonstrated in the example main program \texttt{main341}. 

We have validated the matching strategy with cross-checks against \sherpa\ and \powhegbox\ by comparing the cross sections and $K$-factors from different generators with different event-generation strategies. We observe a very good agreement of the hard-process cross sections throughout the whole kinematic range for all the considered DIS invariants corresponding to a typical HERA setup. 

We have studied the effects of NLO corrections in reduced cross-sections of neutral-current positron-proton DIS and compared the predictions to measurements from ZEUS and H1 collaborations at the HERA collider, yielding greatly reduced scale uncertainties and better overall agreement to data. 
When considering exclusive 1-jet cross sections at the EIC kinematics, we notice that the NLO matching provides again significantly reduced scale variations and reduces dependence on parton-shower implementation thanks to the included matrix-element corrections for the first emission.

The implemented multiplicative matching provides the first fully internal NLO-accurate simulations in \pythia\ for high-energy collisions involving hadron beams. In addition to being very convenient for users to have only a single code to work with, another advantage is that internal kinematics and parton-shower evolution scales are now consistently accounted for throughout the whole event-generation process, without mismatches between evolution scales of the first and subsequent emissions. While the presented implementation considers neutral-current DIS processes, the introduced framework for the $K$-factor calculation and matching procedure can be extended to other scattering processes such as charged-current DIS, vector-boson-fusion Higgs production, and Drell-Yan dilepton production in hadronic collisions.
Extensions to other color-singlet processes, such as gluon-fusion Higgs production or pure QCD processes, may be envisioned in the future as well.
The latter two examples will require work to implement Born-local $K$-factors beyond the structure-function approach, which our generic implementation readily allows for.

\begin{acknowledgements}
The authors would like to thank Silvia Ferrario-Ravasio for answering questions on the implementation of the DIS process in \powhegbox, and Sami Yrjänheikki for providing independent cross-checks on NLO DIS structure functions. 
This research was funded through the Research Council of Finland, projects 331545, 336419 and 361179, and was a part of the Center of Excellence in Quark Matter of the Research Council of Finland, project 346326.
The work of CTP was supported by the Deutsche Forschungsgemeinschaft (DFG) under grant 396021762 - TRR 257: Particle Physics Phenomenology after the Higgs Discovery.
The authors wish to acknowledge CSC – IT Center for Science, Finland, for computational resources. 
\end{acknowledgements}

\appendix

\begin{figure*}
     \centering
     \begin{subfigure}[b]{0.45\textwidth}
         \centering
         \includegraphics[width=\textwidth]{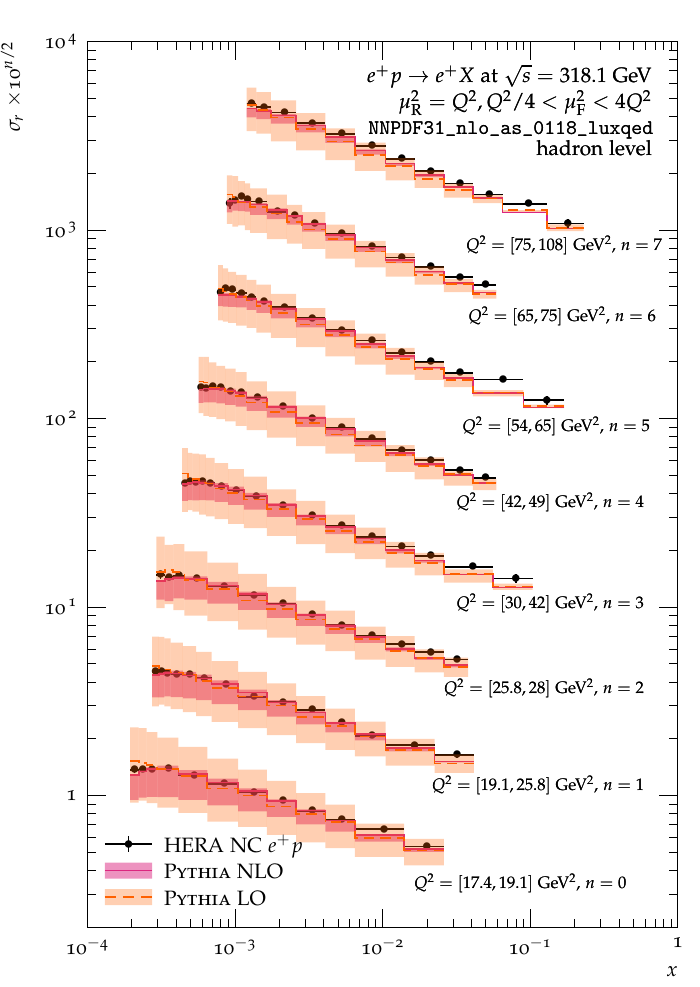}
     \end{subfigure}
     \hfill
     \begin{subfigure}[b]{0.45\textwidth}
         \centering
         \includegraphics[width=\textwidth]{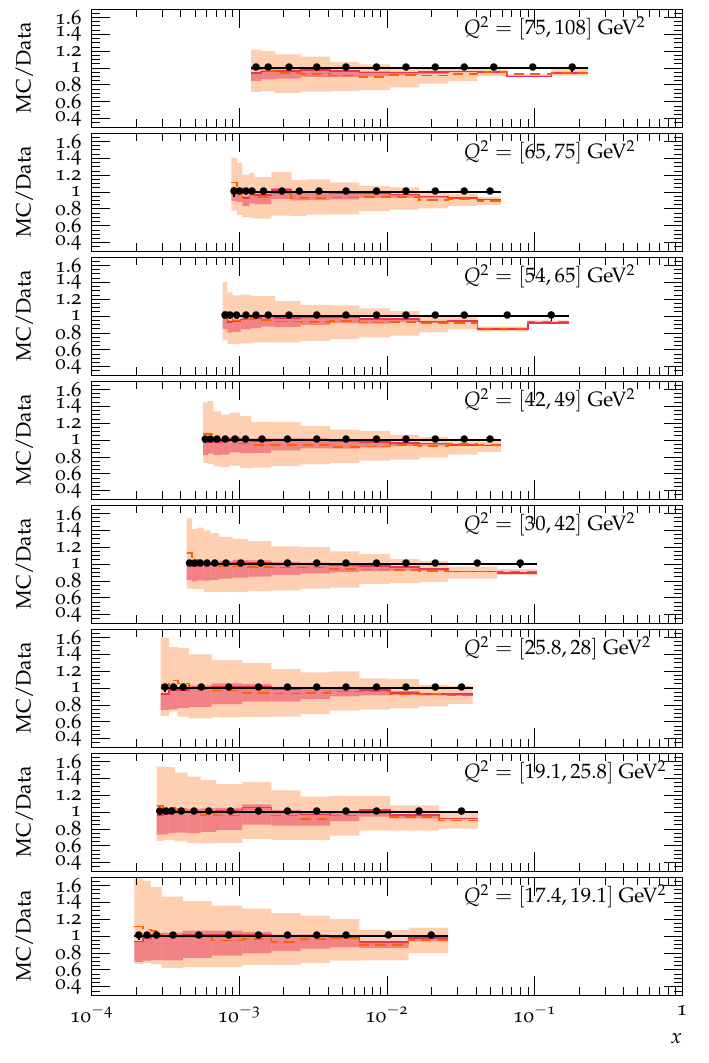}
     \end{subfigure}
        \caption{Reduced cross-sections from $Q^2 = 17.4$ GeV$^2$ to $108$ GeV$^2$. Data from \cite{H1:2015ubc}. }
        \label{fig:HERA2}
\end{figure*}

\begin{figure*}
     \centering
     \begin{subfigure}[b]{0.45\textwidth}
         \centering
         \includegraphics[width=\textwidth]{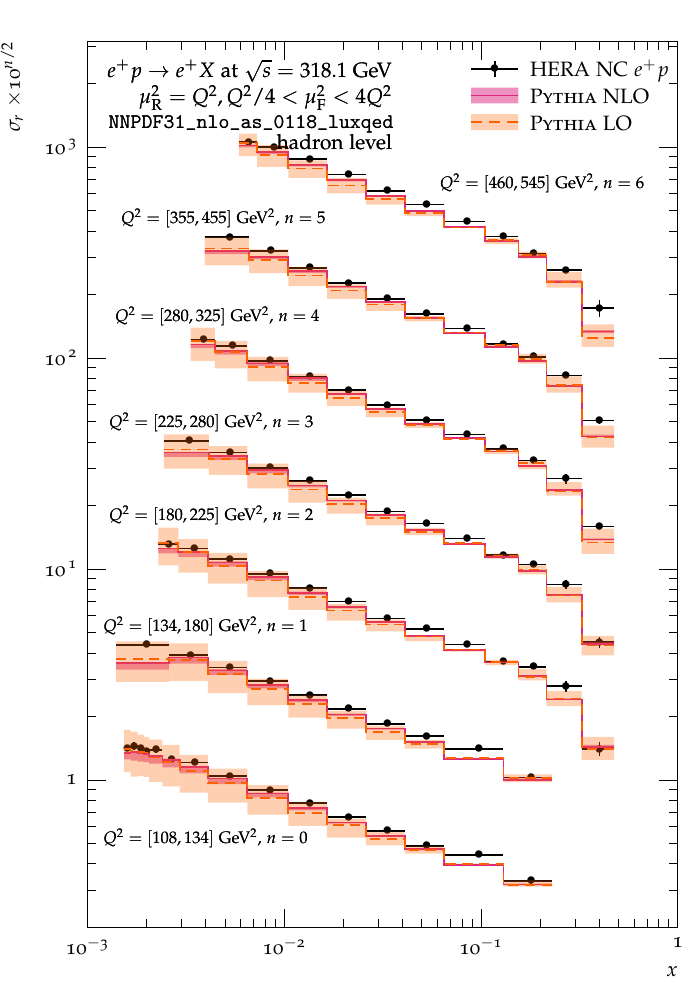}
     \end{subfigure}
     \hfill
     \begin{subfigure}[b]{0.45\textwidth}
         \centering
         \includegraphics[width=\textwidth]{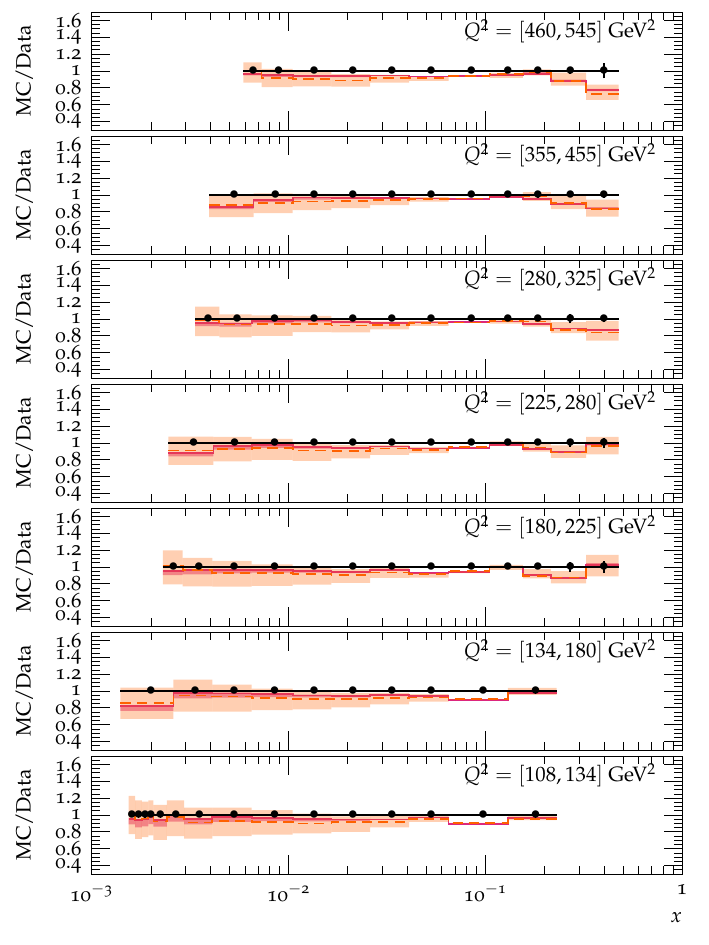}
     \end{subfigure}
        \caption{Reduced cross-sections from $Q^2 = 108$ GeV$^2$ to $545$ GeV$^2$. Data from \cite{H1:2015ubc}. }
        \label{fig:HERA3}
\end{figure*}

\begin{figure*}
     \centering
     \begin{subfigure}[b]{0.45\textwidth}
         \centering
         \includegraphics[width=\textwidth]{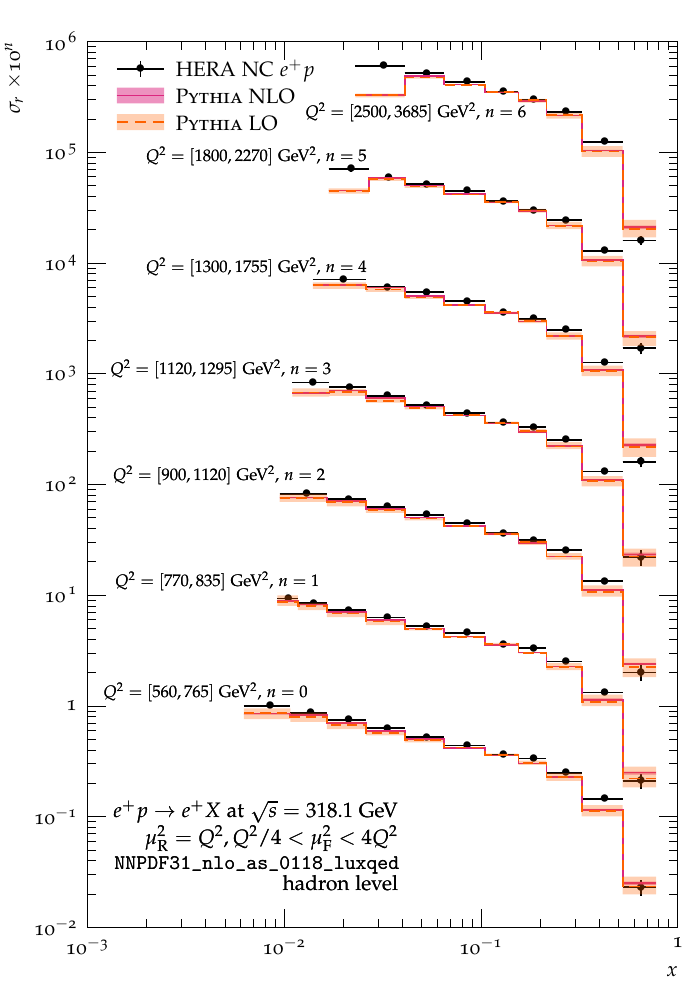}
     \end{subfigure}
     \hfill
     \begin{subfigure}[b]{0.45\textwidth}
         \centering
         \includegraphics[width=\textwidth]{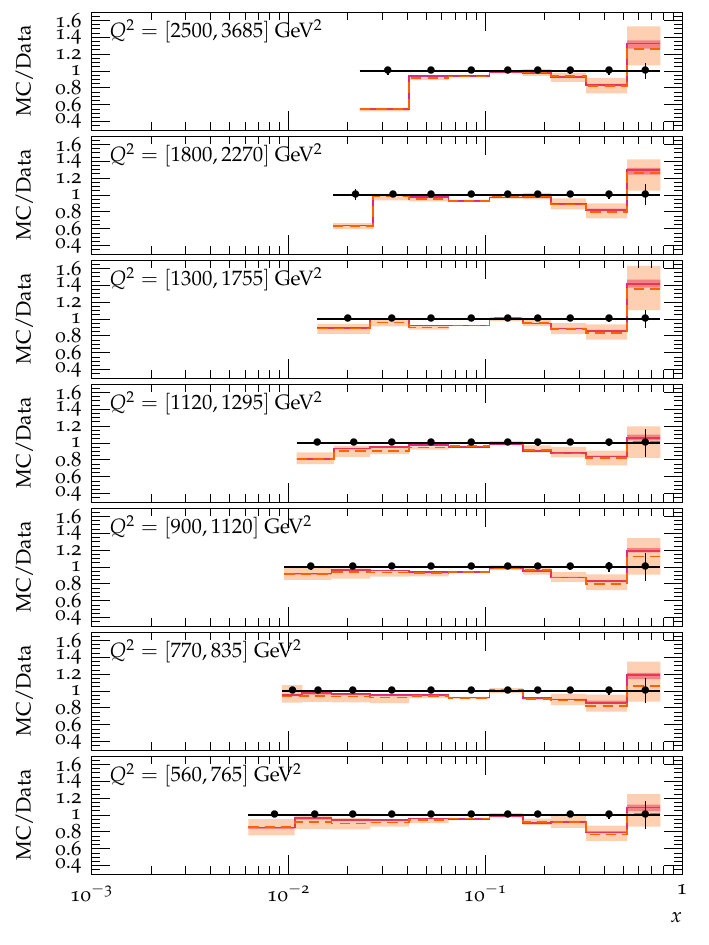}
     \end{subfigure}
        \caption{Reduced cross-sections from $Q^2 = 560$ GeV$^2$ to $3685$ GeV$^2$. Data from \cite{H1:2015ubc}. }
        \label{fig:HERA4}
\end{figure*}

\begin{figure*}
     \centering
     \begin{subfigure}[b]{0.45\textwidth}
         \centering
         \includegraphics[width=\textwidth]{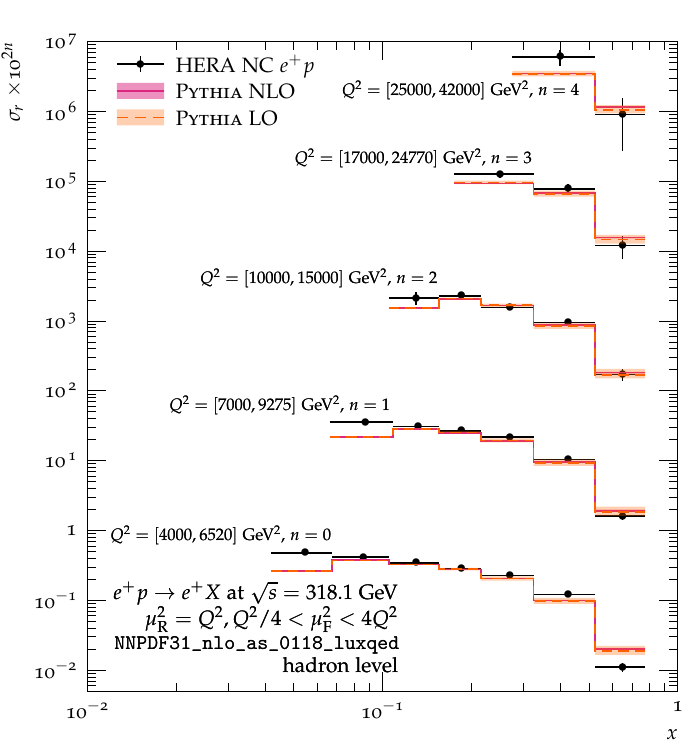}
     \end{subfigure}
     \hfill
     \begin{subfigure}[b]{0.45\textwidth}
         \centering
         \includegraphics[width=\textwidth]{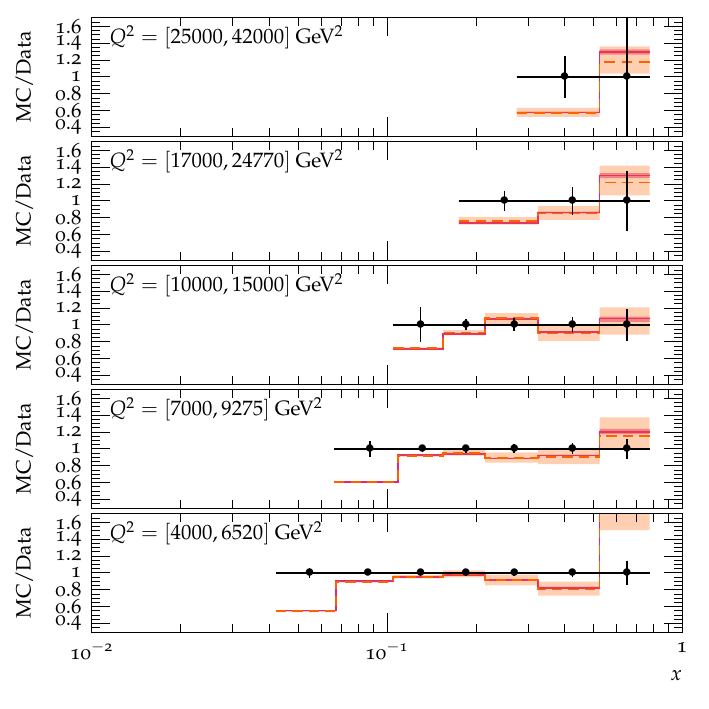}
     \end{subfigure}
        \caption{Reduced cross-sections from $Q^2 = 4000$ GeV$^2$ to $42000$ GeV$^2$. Data from \cite{H1:2015ubc}. }
        \label{fig:HERA5}
\end{figure*}

\section{Comparisons to HERA reduced cross-section measurements}
\label{app}

Figures \ref{fig:HERA2}-\ref{fig:HERA5} show the reduced cross sections from $Q^2 = 17.4$ GeV$^2$ to $Q^2 = 42\,000$ GeV$^2$ as a function of $x$. The simulation undershoots the data at the lowest $x$ bins at high-$Q^2$.
Some of these data bins reside at the edge of the phase space, and in some cases even extend to the kinematically forbidden region. Consider the relation $Q^2 = xsy$ which holds for massless particles. This determines a minimum value $x_\mathrm{min} = Q^2_\mathrm{min} / s$ for a given $Q^2$ bin. The event generation respects this limit, and the reduced cross-sections drop near the phase space edge, while the measured data extends to even lower values in some cases. As a concrete example, the first $x$ bin in the $n=6$ data of Fig.~\ref{fig:HERA4} has a kinematical limit approximately at $x_\mathrm{min} = 0.0247$, which is larger than $x = 0.0230$, the bin edge of the data point.

\clearpage

\bibliographystyle{spphys}
\bibliography{main}

\begin{thebibliography}{10}
\providecommand{\url}[1]{{#1}}
\providecommand{\urlprefix}{URL }
\expandafter\ifx\csname urlstyle\endcsname\relax
  \providecommand{\doi}[1]{DOI \discretionary{}{}{}#1}\else
  \providecommand{\doi}{DOI \discretionary{}{}{}\begingroup
  \urlstyle{rm}\Url}\fi

\bibitem{Buckley:2011ms}
A.~Buckley, et~al., Phys. Rept. \textbf{504}, 145 (2011).
\newblock \doi{10.1016/j.physrep.2011.03.005}

\bibitem{CTEQ:1993hwr}
R.~Brock, et~al., Rev. Mod. Phys. \textbf{67}, 157 (1995).
\newblock \doi{10.1103/RevModPhys.67.157}

\bibitem{Gribov:1972ri}
V.N. Gribov, L.N. Lipatov, Sov. J. Nucl. Phys. \textbf{15}, 438 (1972)

\bibitem{Lipatov:1974qm}
L.N. Lipatov, Yad. Fiz. \textbf{20}, 181 (1974)

\bibitem{Dokshitzer:1977sg}
Y.L. Dokshitzer, Sov. Phys. JETP \textbf{46}, 641 (1977)

\bibitem{Altarelli:1977zs}
G.~Altarelli, G.~Parisi, Nucl. Phys. B \textbf{126}, 298 (1977).
\newblock \doi{10.1016/0550-3213(77)90384-4}

\bibitem{Frixione:2002ik}
S.~Frixione, B.R. Webber, JHEP \textbf{06}, 029 (2002).
\newblock \doi{10.1088/1126-6708/2002/06/029}

\bibitem{Nason:2004rx}
P.~Nason, JHEP \textbf{11}, 040 (2004).
\newblock \doi{10.1088/1126-6708/2004/11/040}

\bibitem{Bengtsson:1986hr}
M.~Bengtsson, T.~Sjöstrand, Phys. Lett. B \textbf{185}, 435 (1987).
\newblock \doi{10.1016/0370-2693(87)91031-8}

\bibitem{Miu:1998ju}
G.~Miu, T.~Sjöstrand, Phys. Lett. B \textbf{449}, 313 (1999).
\newblock \doi{10.1016/S0370-2693(99)00068-4}

\bibitem{Hoeche:2011fd}
S.~Hoeche, F.~Krauss, M.~Schonherr, F.~Siegert, JHEP \textbf{09}, 049 (2012).
\newblock \doi{10.1007/JHEP09(2012)049}

\bibitem{Frixione:2007vw}
S.~Frixione, P.~Nason, C.~Oleari, JHEP \textbf{11}, 070 (2007).
\newblock \doi{10.1088/1126-6708/2007/11/070}

\bibitem{Alioli:2010xd}
S.~Alioli, P.~Nason, C.~Oleari, E.~Re, JHEP \textbf{06}, 043 (2010).
\newblock \doi{10.1007/JHEP06(2010)043}

\bibitem{Hoche:2010pf}
S.~Hoche, F.~Krauss, M.~Schonherr, F.~Siegert, JHEP \textbf{04}, 024 (2011).
\newblock \doi{10.1007/JHEP04(2011)024}

\bibitem{Platzer:2011bc}
S.~Platzer, S.~Gieseke, Eur. Phys. J. C \textbf{72}, 2187 (2012).
\newblock \doi{10.1140/epjc/s10052-012-2187-7}

\bibitem{Alwall:2014hca}
J.~Alwall, R.~Frederix, S.~Frixione, V.~Hirschi, F.~Maltoni, O.~Mattelaer, H.S.
  Shao, T.~Stelzer, P.~Torrielli, M.~Zaro, JHEP \textbf{07}, 079 (2014).
\newblock \doi{10.1007/JHEP07(2014)079}

\bibitem{AbdulKhalek:2021gbh}
R.~Abdul~Khalek, et~al., Nucl. Phys. A \textbf{1026}, 122447 (2022).
\newblock \doi{10.1016/j.nuclphysa.2022.122447}

\bibitem{Banfi:2023mhz}
A.~Banfi, S.~Ferrario~Ravasio, B.~J\"ager, A.~Karlberg, F.~Reichenbach,
  G.~Zanderighi, JHEP \textbf{02}, 023 (2024).
\newblock \doi{10.1007/JHEP02(2024)023}

\bibitem{Jezo:2015aia}
T.~Je{\v{z}}o, P.~Nason, JHEP \textbf{12}, 065 (2015).
\newblock \doi{10.1007/JHEP12(2015)065}

\bibitem{Borsa:2024rmh}
I.~Borsa, B.~J\"ager, JHEP \textbf{07}, 177 (2024).
\newblock \doi{10.1007/JHEP07(2024)177}

\bibitem{Jadach:2011cr}
S.~Jadach, A.~Kusina, W.~Placzek, M.~Skrzypek, M.~Slawinska, Phys. Rev. D
  \textbf{87}(3), 034029 (2013).
\newblock \doi{10.1103/PhysRevD.87.034029}

\bibitem{Bellm:2025pcw}
J.~Bellm, et~al., arXiv:2512.16645  (2025)

\bibitem{DErrico:2011wfa}
L.~D'Errico, P.~Richardson, Eur. Phys. J. C \textbf{72}, 2042 (2012).
\newblock \doi{10.1140/epjc/s10052-012-2042-x}

\bibitem{Masouminia:2026bhk}
M.R. Masouminia, A.~Papaefstathiou, P.~Richardson,   (2026)

\bibitem{vanBeekveld:2023chs}
M.~van Beekveld, S.~Ferrario~Ravasio, JHEP \textbf{02}, 001 (2024).
\newblock \doi{10.1007/JHEP02(2024)001}

\bibitem{vanBeekveld:2025lpz}
M.~van Beekveld, S.~Ferrario~Ravasio, J.~Helliwell, A.~Karlberg, G.P. Salam,
  L.~Scyboz, A.~Soto-Ontoso, G.~Soyez, S.~Zanoli, JHEP \textbf{10}, 038 (2025).
\newblock \doi{10.1007/JHEP10(2025)038}

\bibitem{Cacciari:2015jma}
M.~Cacciari, F.A. Dreyer, A.~Karlberg, G.P. Salam, G.~Zanderighi, Phys. Rev.
  Lett. \textbf{115}(8), 082002 (2015).
\newblock \doi{10.1103/PhysRevLett.115.082002}.
\newblock [Erratum: Phys.Rev.Lett. 120, 139901 (2018)]

\bibitem{Currie:2018fgr}
J.~Currie, T.~Gehrmann, E.W.N. Glover, A.~Huss, J.~Niehues, A.~Vogt, JHEP
  \textbf{05}, 209 (2018).
\newblock \doi{10.1007/JHEP05(2018)209}

\bibitem{Gehrmann:2018odt}
T.~Gehrmann, A.~Huss, J.~Niehues, A.~Vogt, D.M. Walker, Phys. Lett. B
  \textbf{792}, 182 (2019).
\newblock \doi{10.1016/j.physletb.2019.03.003}

\bibitem{NNLOJET:2025rno}
A.~Huss, et~al., SciPost Phys. Codeb. \textbf{69}, 1 (2026).
\newblock \doi{10.21468/SciPostPhysCodeb.69}

\bibitem{Karlberg:2024hnl}
A.~Karlberg, SciPost Phys. Codeb. \textbf{2024}, 32 (2024).
\newblock \doi{10.21468/SciPostPhysCodeb.32}

\bibitem{Borsa:2022cap}
I.~Borsa, D.~de~Florian, I.~Pedron, Phys. Rev. D \textbf{107}(5), 054027
  (2023).
\newblock \doi{10.1103/PhysRevD.107.054027}

\bibitem{Sherpa:2024mfk}
E.~Bothmann, et~al., JHEP \textbf{12}, 156 (2024).
\newblock \doi{10.1007/JHEP12(2024)156}

\bibitem{Hoche:2018gti}
S.~H{\"o}che, S.~Kuttimalai, Y.~Li, Phys. Rev. D \textbf{98}(11), 114013
  (2018).
\newblock \doi{10.1103/PhysRevD.98.114013}

\bibitem{Meinzinger:2025pam}
P.~Meinzinger, D.~Reichelt, F.~Silvetti, Phys. Rev. D \textbf{112}(7), 074039
  (2025).
\newblock \doi{10.1103/1c38-jrb1}

\bibitem{Bierlich:2022pfr}
C.~Bierlich, et~al., SciPost Phys. Codeb. \textbf{2022}, 8 (2022).
\newblock \doi{10.21468/SciPostPhysCodeb.8}

\bibitem{Cabouat:2017rzi}
B.~Cabouat, T.~Sj{\"o}strand, Eur. Phys. J. C \textbf{78}(3), 226 (2018).
\newblock \doi{10.1140/epjc/s10052-018-5645-z}

\bibitem{Helenius:2024wjg}
I.~Helenius, J.O. Laulainen, C.T. Preuss, JHEP \textbf{05}, 153 (2025).
\newblock \doi{10.1007/JHEP05(2025)153}

\bibitem{Bengtsson:1986et}
M.~Bengtsson, T.~Sjöstrand, Nucl. Phys. B \textbf{289}, 810 (1987).
\newblock \doi{10.1016/0550-3213(87)90407-X}

\bibitem{Preuss:2024vyu}
C.T. Preuss, JHEP \textbf{07}, 161 (2024).
\newblock \doi{10.1007/JHEP07(2024)161}

\bibitem{Altarelli:1979ub}
G.~Altarelli, R.K. Ellis, G.~Martinelli, Nucl. Phys. B \textbf{157}, 461
  (1979).
\newblock \doi{10.1016/0550-3213(79)90116-0}

\bibitem{Furmanski:1981cw}
W.~Furmanski, R.~Petronzio, Z. Phys. C \textbf{11}, 293 (1982).
\newblock \doi{10.1007/BF01578280}

\bibitem{Gluck:1994uf}
M.~Gluck, E.~Reya, A.~Vogt, Z. Phys. C \textbf{67}, 433 (1995).
\newblock \doi{10.1007/BF01624586}

\bibitem{Moch:1999eb}
S.~Moch, J.A.M. Vermaseren, Nucl. Phys. B \textbf{573}, 853 (2000).
\newblock \doi{10.1016/S0550-3213(00)00045-6}

\bibitem{Daleo:2006xa}
A.~Daleo, T.~Gehrmann, D.~Maitre, JHEP \textbf{04}, 016 (2007).
\newblock \doi{10.1088/1126-6708/2007/04/016}

\bibitem{ParticleDataGroup:2024cfk}
S.~Navas, et~al., Phys. Rev. D \textbf{110}(3), 030001 (2024).
\newblock \doi{10.1103/PhysRevD.110.030001}

\bibitem{Collins:1989gx}
J.C. Collins, D.E. Soper, G.F. Sterman, Adv. Ser. Direct. High Energy Phys.
  \textbf{5}, 1 (1989).
\newblock \doi{10.1142/9789814503266_0001}

\bibitem{Catani:1996vz}
S.~Catani, M.H. Seymour, Nucl. Phys. B \textbf{485}, 291 (1997).
\newblock \doi{10.1016/S0550-3213(96)00589-5}.
\newblock [Erratum: Nucl.Phys.B 510, 503--504 (1998)]

\bibitem{Sjostrand:2004ef}
T.~Sjöstrand, P.Z. Skands, Eur. Phys. J. C \textbf{39}, 129 (2005).
\newblock \doi{10.1140/epjc/s2004-02084-y}

\bibitem{Brooks:2020upa}
H.~Brooks, C.T. Preuss, P.~Skands, JHEP \textbf{07}, 032 (2020).
\newblock \doi{10.1007/JHEP07(2020)032}

\bibitem{Lonnblad:1992tz}
L.~Lönnblad, Comput. Phys. Commun. \textbf{71}, 15 (1992).
\newblock \doi{10.1016/0010-4655(92)90068-A}

\bibitem{Plehn:2005cq}
T.~Plehn, D.~Rainwater, P.Z. Skands, Phys. Lett. B \textbf{645}, 217 (2007).
\newblock \doi{10.1016/j.physletb.2006.12.009}

\bibitem{Schumann:2007mg}
S.~Schumann, F.~Krauss, JHEP \textbf{03}, 038 (2008).
\newblock \doi{10.1088/1126-6708/2008/03/038}

\bibitem{Krauss:2001iv}
F.~Krauss, R.~Kuhn, G.~Soff, JHEP \textbf{02}, 044 (2002).
\newblock \doi{10.1088/1126-6708/2002/02/044}

\bibitem{Gleisberg:2008fv}
T.~Gleisberg, S.~Hoeche, JHEP \textbf{12}, 039 (2008).
\newblock \doi{10.1088/1126-6708/2008/12/039}

\bibitem{Gleisberg:2007md}
T.~Gleisberg, F.~Krauss, Eur. Phys. J. C \textbf{53}, 501 (2008).
\newblock \doi{10.1140/epjc/s10052-007-0495-0}

\bibitem{Chahal:2022rid}
G.S. Chahal, F.~Krauss, SciPost Phys. \textbf{13}(2), 019 (2022).
\newblock \doi{10.21468/SciPostPhys.13.2.019}

\bibitem{Bothmann:2016nao}
E.~Bothmann, M.~Sch{\"o}nherr, S.~Schumann, Eur. Phys. J. C \textbf{76}(11),
  590 (2016).
\newblock \doi{10.1140/epjc/s10052-016-4430-0}

\bibitem{NNPDF:2017mvq}
R.D. Ball, et~al., Eur. Phys. J. C \textbf{77}(10), 663 (2017).
\newblock \doi{10.1140/epjc/s10052-017-5199-5}

\bibitem{Buckley:2014ana}
A.~Buckley, J.~Ferrando, S.~Lloyd, K.~Nordstr{\"o}m, B.~Page, M.~R{\"u}fenacht,
  M.~Sch{\"o}nherr, G.~Watt, Eur. Phys. J. C \textbf{75}, 132 (2015).
\newblock \doi{10.1140/epjc/s10052-015-3318-8}

\bibitem{H1:2015ubc}
H.~Abramowicz, et~al., Eur. Phys. J. C \textbf{75}(12), 580 (2015).
\newblock \doi{10.1140/epjc/s10052-015-3710-4}

\bibitem{Bierlich:2019rhm}
C.~Bierlich, et~al., SciPost Phys. \textbf{8}, 026 (2020).
\newblock \doi{10.21468/SciPostPhys.8.2.026}

\bibitem{Cacciari:2011ma}
M.~Cacciari, G.P. Salam, G.~Soyez, Eur. Phys. J. C \textbf{72}, 1896 (2012).
\newblock \doi{10.1140/epjc/s10052-012-1896-2}

\bibitem{Carli:2010cg}
T.~Carli, T.~Gehrmann, S.~Hoeche, Eur. Phys. J. C \textbf{67}, 73 (2010).
\newblock \doi{10.1140/epjc/s10052-010-1261-2}

\end{thebibliography}

\end{document}